%% file: onetask_main.tex
\documentclass[10pt,sigconf,letterpaper,nonacm]{acmart}

\usepackage{multirow}
\usepackage{threeparttable}
\usepackage{graphicx}
\usepackage{subcaption}

\usepackage{caption}
\usepackage{booktabs}
\usepackage{array} 
\usepackage{tikz}

\AtBeginDocument{%
  }

\setcopyright{acmlicensed}
\copyrightyear{2025}
\acmYear{2025}
\acmDOI{XXXXXXX.XXXXXXX}
\acmConference[Conference acronym 'XX]{Make sure to enter the correct
  conference title from your rights confirmation email}{June 03--05,
  2018}{Woodstock, NY}
\acmISBN{978-1-4503-XXXX-X/2018/06}

\begin{document}

\title{One Task to Rule Them All: A Closer Look at Traffic Classification Generalizability}

\author[Elham Akbari, 
    Zihao Zhou, 
    M. A. Salahuddin, Noura Limam et al.]{Elham Akbari, 
    Zihao Zhou, 
    M. A. Salahuddin,
 	 Noura Limam, Raouf Boutaba$^{*}$, \\Bertrand Mathieu, Stephanie Moteau, and Stephane Tuffin$^{\dagger}$}

\affiliation{%
  \institution{$^{*}$University of Waterloo, Ontario, Canada\quad
               $^{\dagger}$Orange Labs, Lannion, France \quad}
  \country{}
}



\begin{abstract} 

Existing website fingerprinting and traffic classification solutions do not work well when the evaluation context changes, as their performances often heavily rely on context-specific assumptions. To clarify this problem, we take three prior solutions presented for different but similar traffic classification and website fingerprinting tasks, and apply each solution's model to another solution's dataset. We pinpoint dataset-specific and model-specific properties that lead each of them to overperform in their specific evaluation context.

As a realistic evaluation context that takes practical labeling constraints into account, we design an evaluation framework using two recent real-world TLS traffic datasets from large-scale networks. The framework simulates a futuristic scenario in which SNIs are hidden in some networks but not in others, and the classifier's goal is to predict destination services in one network's traffic, having been trained on a labelled dataset collected from a different network. Our framework has the distinction of including real-world distribution shift, while excluding concept drift. We show that, even when abundant labeled data is available, the best solutions' performances under distribution shift are between 30\% and 40\%, and a simple 1-Nearest Neighbor classifier's performance is not far behind. We depict all performances measured on different models, not just the best ones, for a fair representation of traffic models in practice.


\end{abstract}




\newcommand{\etal}{et al. }
\newcommand{\cf}{cf., }
\newcommand{\eg}{e.g., }
\newcommand{\ie}{i.e., }
\newcommand{\etc}{etc. }
\newcommand{\wrt}{w.r.t. }
\newcommand{\aka}{a.k.a. }
\newcommand{\red}[1]{{\color{red}#1}}
\newcommand{\blue}[1]{{\color{blue}#1}}

\newcommand{\diagonalSplitCell}[2]{%
  \begin{tikzpicture}[baseline=(current bounding box.center)]
    \draw (0,0) rectangle (2.8,1.2);         
    \draw (0,1.2) -- (2.8,0);                
    \node[anchor=north west, font=\scriptsize, align=left] at (0.1,1.15) {#1}; 
    \node[anchor=south east, font=\scriptsize, align=right] at (2.7,0.05) {#2}; 
  \end{tikzpicture}%
}

\settopmatter{printfolios=true}

\maketitle

\input{new_sections/intro2}

\input{new_sections/relatedwork}

\input{new_sections/on-generalizability}

\input{new_sections/revisiting-prior}

\input{new_sections/prior-performance}
\input{new_sections/prior-performance-contd}

\input{new_sections/model-generalizability-main}
\input{new_sections/gen-experiments}

\input{new_sections/conclusion}

\newpage
\bibliographystyle{ACM-Reference-Format}
\bibliography{bibliography}

\appendix

\section{Ethics}
The datasets used in this work were all publicly available, except the Orange datasets. The Orange datasets were used in compliance with the legal frameworks in force in the countries where they were collected, such as the EU General Data Protection Regulation (GDPR) and national laws resulting from the EU Digital Agenda. A specific ``Personal Data Protection'' policy has been defined for the data collection group and is applied both internally and with all the partners. The data was anonymized by the collection group before being shared by partners.

\input{new_sections/appendix1}

\input{new_sections/appendix-model-listing}

\end{document}

%% file: new_sections/intro2.tex
\section{Introduction}

An increasing demand for user privacy has led to a growing adoption of encrypted protocols, notably TLS and QUIC, by web and mobile platforms to protect users' data from man-in-the-middle and man-on-the-side attacks. Encrypted protocols ensure data privacy, but do not attempt to hide the destination server name and allow its transmission in clear text by the \texttt{ClientHello} message in the Server Name Information (SNI) field in TLS, and a similar field in QUIC. 



Despite attempts to push the SNI in the encrypted channel in the original plan for TLS 1.3, clear-text domain names are widely present in today's TLS traffic and used by network service providers as the de facto basis for traffic classification. 
However, concerns exist that if SNI encryption (ESNI) is widely adopted, it can bring the ISP systems that depend on traffic classification to a halt.  

 
The researcher's view of the story is somewhat different. While the widespread adoption of ESNI and Encrypted Client Hello undermines fast and large-scale traffic classification in practice, attacks can still be conducted on traffic to reveal its destination with high accuracy, as is the case in numerous Tor website fingerprinting and encrypted traffic classification works \cite{ rossilessonslearned, aceto2020toward, iman, flowpic, flowprint2020, taylor2016appscanner, etbert}. Depending on each work's specific task, these works propose methods to identify the website, application or destination service associated with Tor, VPN, TLS or QUIC traffic and often report very high accuracy. 

Looking at these works' results, it is easy to conclude that ESNI identification is a resolved research problem. 
For example, \cite{etbert}'s proposed Large Language Model (LLM) is reported to achieve close to 100\% accuracy in application and service identification, by only receiving a single encrypted packet as input. As another example, \cite{dfattack}'s proposed method is reported to achieve near perfect accuracy on undefended Tor traffic by only looking at its sequence of packet directions. 

However, the devil is in the details, and a closer look at the literature raises questions on the meaning of these results. For example, DFattack \cite{dfattack} reached 97\% accuracy in solving a 95-way traffic classification task on Tor traffic and only needed packet directions to do so. On the other hand, other works \cite{iman, pescape2021xaimirage} had to propose a multimodal deep model (operating on many more features than mere packet directions) to solve an easier task with fewer classes compared to DFattack's task. The target traffic in \cite{iman, pescape2021xaimirage} is TLS, not Tor, but does that mean that Tor fingerprinting is easier than TLS fingerprinting? As another example, ET-BERT \cite{etbert} only relies on a single packet's bytes to identify its category with over 97\% accuracy. Other works \cite{pescape2021xaimirage, nprint}, although admittedly published slightly earlier, have found similar classification tasks significantly more challenging; \cite{pescape2021xaimirage} used the packet bytes of an entire flow to solve an identical task to \cite{etbert}'s on a different dataset, and nPrint \cite{nprint} proposed an elaborate representation of the same input as \cite{etbert} for an OS fingerprinting task. There seems to be a motivation that drives the suggestion of new classifiers for identical and related tasks, even though the previous ones are perfect as far as their authors' evaluations suggest.



The motivation is, simply put, lack of generalizability to very close \textit{task}s (defined in Section \ref{sec:definitions}) and datasets. Proposed traffic models often do not work well when presented to a new dataset, and finding the reason is not always straightforward. The models are often unexplainable, so the reasons for their failure are as obscure as the reasons for their success. In the absence of explainability, the only way to assess a model's performance is by performance metrics, which can wildly differ depending on evaluation process choices, \eg train-to-test split ratio, data selection criteria such as cutoff points, selection size, number of classes, etc. 

Since evaluation choices have a wide range, reasons for over-evaluation need to be addressed case by case, as was done in \cite{dosanddonts22}. We perform a similar study in Section \ref{sec:dataset-specificity} on three \emph{time-series} traffic classifiers. Time-series models, in our view, are more suitable for tasks involving large-scale networks, such as ESNI identification, than models relying on packet bytes, because of the high cost of collecting and storing packet bytes that can be trusted as representative of a large network. The three works \cite{dfattack, flowpic, elham_cnsm} were selected based on the extensive lines of research that either led to them or were built on them \cite{tiktok2019, flowpic2023replication, miniflowpic, iman, navid2022datadrift}. Each work is representative of a different type of deep model: DFattack \cite{dfattack} is a well-established 1-dimensional Convolutional Neural Network (1D-CNN), Flowpic \cite{flowpic} is a 2D-CNN with an intuitive feature representation method, and UWTransformer \cite{elham_cnsm} is a Transformer-based model, shown to outperform the authors' previous time-series models. In Section \ref{sec:dataset-specificity}, we compare the performance of the three methods on one evaluation dataset from each work, uncovering a pattern of reproducible but \textit{dataset-specific} performances.

While the evaluation tasks in Section \ref{sec:dataset-specificity} are similar, they are not identical, and do not rule out the possibility that each model may be task-specific rather than dataset-specific. 
Hence, Sections \ref{sec:model-gen-main}-\ref{sec:model-gen-shift} focus on one task, ESNI identification in TLS traffic and introduce an evaluation framework based on two different real-world datasets representing the exact same task in every detail found to be important in the previous section. The evaluation framework has the distinction of simulating a realistic ESNI identification scenario by taking into account labeling constraints in the real world. Up to 19 traffic classifiers, including classical and deep models, are evaluated on it in Section \ref{sec:model-gen-shift}. Fifteen of these models are issued by hyperparameter tuning on a Transformer-based model, which we argue can also be viewed as a generic case of 1D-CNNs. Although hyperparameter tuning may sound trivial, it is not so when the scope of hyperparameters is large. From one point of view, any proposed method can be viewed as only a set of suggested hyperparameter values.

Our contributions are:
\begin{enumerate}
    \item Bringing together TLS traffic classification and Tor website fingerprinting tasks by comparing two models and two datasets proposed for each, and showing that their models have comparable performances on different datasets regardless of task, as long as the model was not evaluated on the dataset in its original paper. In the latter case, the model's performance on the dataset is unusually good.
    
    \item Developing a model evaluation framework that involves realistic distribution shift for an ESNI identification task, while it controls for concept drift. Showing that distribution shift significantly degrades model performance even when concept drift is not present.

    \item Demonstrating that over-parameterized deep classifiers show better generalizability than their smaller counterparts, while also showing that differences in \textit{best} generalizable performances are small across models, and notably smaller than differences in performance readings from the same model when trained by distinct processes until convergence.

    \item Putting existing traffic classifiers in perspective by showing their sensitivity to non-representative dataset characteristics or best hyperparameter values which are not guaranteed under realistic assumptions, and demonstrating their average-case performance based on multiple datasets.
    
\end{enumerate}

The paper is organized as follows. Section \ref{sec:relatedwork} describes the related work. Section \ref{sec:definitions} defines key terms used in the rest of the paper, such as generalizability, task, representativeness, methods, and models. Sections \ref{sec:dataset-specificity}-\ref{sec:model-gen-shift} include our evaluations and results and Section \ref{sec:conclusion} concludes the paper. Our work's artifacts, including data preparation for the test framework's public dataset and model building and training scripts can be found in
\hyperlink{https://github.com/one-task-traffic/traffic-generalization-study}{https://github.com/one-task-traffic/traffic-generalization-study}.

%% file: new_sections/relatedwork.tex
\section{Related Works}\label{sec:relatedwork}
Encrypted traffic classification is a mature field of research with various aliases in different works, \eg website and application fingerprinting \cite{wf2009herrmann, taylor2016appscanner, wftriplet, wfattacks2023, appfingerprinting2022}, OS fingerprinting \cite{nprint}, protocol identification \cite{ggfast}, streaming application identification, malware detection \cite{netfound23}, and quality of service inference \cite{bronzino2019inferring}, to name a few. It was originally tackled using classical ML models, such as Naive Bayes \cite{wf2009herrmann}, kNN \cite{wfrealistictorattack2016}, and tree-based methods \cite{randomforestforTC2017} in the earlier days, and using deep learning since 2017  \cite{aceto2019mimetic, iman, gnn2023tnsm, etbert, dfattack}. \citet{rossilessonslearned} give an apt account of successive waves of ML models and their adoption for traffic classification. 

Among ML models, one-dimensional Convolutional Neural Networks (1D-CNNs)\cite{wang20171dcnn} and XGBoost \cite{xgboost2016first} are widely believed to perform well \cite{aceto2019DL, rossilessonslearned}. Despite the hype of deep learning (DL), Classical ML models still appear in more recent works \cite{ggfast, nprint, bronzino2019inferring, rossilessonslearned}, as they have shown comparable performance to DL \cite{aceto2019DL}, work better out of the box due to fewer hyperparameters, and are typically faster to train and evaluate. However, to the best of our knowledge, a systematic study of the \textit{generalizability} of classical vs deep models in traffic classification has not been conducted. We believe this is due to the difficulty of obtaining sufficiently differing datasets that represent the exact same task, which is necessary for a study of generalizability. 

Poor traffic classifier generalizability was observed as early as 2009 \cite{wf2009herrmann, taylor2016appscanner, wfcriticaleval2014, wfrealistictorattack2016, navid2022datadrift}. \cite{taylor2016appscanner} found their trained model to perform poorly on newly collected mobile application traffic and attributed it to a change in application versions. \cite{wf2009herrmann} had previously observed a similar problem and attributed it to \textit{concept drift}.  Similarly, \cite{wfcriticaleval2014} found Tor website fingerprinting models to perform very poorly on traffic from different Tor browser versions, or the same browser version but at a different time. Traffic collection time showed a similar effect in \cite{navid2022datadrift}, degrading the performance of a TLS traffic classifier. The consensus was that changes in traffic render traffic classifiers obsolete in time, and frequent retraining in necessary to counter the effect. 

Other lines of research, however, raise questions on the extent of concept drift's role in performance degradation, and suggest that the degradation may be the result of inflated performance evaluations when training and test instances come from the same dataset. For example, few-shot classifiers perform well on instances of classes unseen in training \cite{guo2022ifip, wftriplet, miniflowpic} -- they perform well on zero-day tasks, but only if the instances come from a dataset very similar to the training dataset \cite{elham_cnsm}. More concerningly, \cite{dosanddonts22, trustee} found fundamental problems with the methodology of a wide range of notable prior works, which can lead to inflated performance evaluations in their proposed traffic classifiers. \cite{dosanddonts22} found instances of test data snooping in the ML pipeline. \cite{trustee} found various instances of \textit{spurious features} in the CIC datasets \cite{iscx-vpn, iscx-tor}, widely used as benchmark in encrypted traffic classification. 

A subsequent attempt by \cite{trustee}'s researchers in \cite{netunicorn_ccs2023} to collect a malware detection dataset revealed the difficulties of collecting a shortcut-free dataset. Only after training a model on the collected data, explaining the model using an explainability tool, comparing the learned knowledge with what the model is supposed to learn, and going back three times to recollect data, were they able to pronounce the data shortcut-free. If what the model should learn is not clear, as is the case in website fingerprinting for example, we believe all datasets should be expected to contain spurious features. Hence, a high performance in same-dataset training and evaluation of a model does not indicate that the model can learn the underlying task. These findings served as a motivation for this work's elaborate test framework (\cf Section \ref{sec:test-framework}).

\textit{Generalizability in ML.}
Evidence of shortcut learning by DL and Large Language Models (LLM) was found by \cite{ml_underspecification22, nature_gen20} in language processing and vision. \cite{ml_underspecification22} highlights the importance of small choices in the training process, \eg initial weights, in the resulting LLM's generalizability, which led to the authors suggesting that generalizability should be seen as a property of a trained model rather than a model type or training pipeline. \cite{nature_gen20} suggested more carefully curated tests as a necessary step towards improving model generalizability. Both \cite{ml_underspecification22} and \cite{nature_gen20} can be viewed as works that describe and provide insights on the problem, but do not really offer a solution.

Model generalizability and robustness to distribution shift is an active field of research  \cite{ generalization2019neyshabur, neyshabur2020transfer, neyshabur2020sharpness, neyshabur2022unlabeled-ood}, with close ties to transfer learning \cite{bendavid2010domainadapt}. 
These works' findings remain somewhat theoretical by our assessment, due to unverifiable assumptions about the input distribution. However, they highlight the role of optimizers and over-parameterization in generalizability. 

Certain findings of the above works are confirmed in this work. Over-parameterized deep models, \ie models with more parameters than necessary to fit the training data perfectly, are believed to be more generalizable than the rest \cite{generalization2019neyshabur}. While the larger models in this work did not fit the data perfectly, we did find improved generalizability in models with an unnecessarily large number of parameters. \cite{neyshabur2020sharpness} describes the loss landscapes of today's deep models as sharp, and therefore the choice of optimizer, which determines the model's loss navigation strategy, important for model generalizability. We found evidence of this when training our designed transformers, but found tuning learning rate schedules more effective than changing the optimizer from Adam, the de facto optimizer in most traffic classifiers.

\textit{Time series Transformers.} The proposed Transformer-based traffic classifiers in the literature \cite{etbert, kdd22flowformer, trafficformer2024sp} mostly take raw or processed traffic packet bytes as input. In contrast, this work explores a \textit{time-series} Transformer-based traffic classifier, which was first proposed in \cite{elham_cnsm}, because packet-byte traffic datasets have been shown to be prone to include shortcuts in \cite{navid2022datadrift, trustee}. Employing Transformers on time-series input is an active and young line of research \cite{worth64words2022, zeng2023transformers, samformer2024}. While \cite{worth64words2022} found them to perform well, \cite{zeng2023transformers} questioned those findings by suggesting a simple linear model that outperforms them. However, recently, \cite{samformer2024} cited that they indeed work well if their training instability is addressed. In line with \cite{samformer2024}, this work also finds training instability to be an important factor in the performance of \cite{elham_cnsm}'s model. To the best of our knowledge, \cite{elham_cnsm}'s model is the only Transformer-based model in traffic classification suggested for time-series input.

%% file: new_sections/on-generalizability.tex
\section{Tasks, datasets, and generalizability} \label{sec:definitions}

In machine learning, model generalizability is defined as the likelihood of a trained ML model performing within a small threshold of its validation performance, when evaluated on a test dataset. 
From a theoretical standpoint, if no assumption is made on the test dataset, ensuring generalizability is as complex as obtaining a model for the new dataset from scratch \cite{bendavid2010domainadapt}. In other words, the existence of a trained classifier does not make a difference to solving the problem posed by a new dataset, if the dataset is not somehow similar to the classifier's validation dataset. 

A common assumption is the identically and independently distributed (i.i.d.) assumption. It states that all data items in the datasets should be identically and independently sampled from the same underlying distribution. While the assumption makes theoretical analysis easier, it is an ideal case and difficult to verify in practice, as the underlying probability distribution is often unknown. Nevertheless, ML models developed under this assumption are shown to work well in practice, in situations where the assumption is believed not to hold.

\textbf{Definition of Task and Dataset}. Even though the underlying distribution is unknown, defining it allows a way to define common terms associated with machine learning. The distribution, denoted as $p(X,y)$, is defined as a joint probability distribution over inputs and labels, and is the result of a real-world random process. In the context of network traffic, this random process is determined by users, devices, software stacks, and the network system. In the same context, each input $x_i$ is a traffic object, represented as a vector, and each label $y_i$, represented by a scalar. The joint distribution is defined over a set of  $x_i$'s, denoted as $X$, and a set of $y_i$'s, denoted as $y$, of possible labels. These sets are defined by the \emph{task}. 

We define a task as a query that can be fully specified by $p(X,y)$. In other words, $p(X,y)$ holds the complete answer to the query specified by the task. The point of an ML model is to predict the probability distribution, denoted as $Prob(y|x)$, of labels when given an input in the domain of the task. $Prob(y|x)$ is learned from the \textit{dataset}. A dataset is defined as a set of $(x_i,y_i)$ tuples sampled from $p(X,y)$. A dataset should ideally be \textit{representative} of a task. 

Let's see how these theoretical definitions relate to practical traffic classification. Consider the following Tor website fingerprinting task as a concrete example: ``Given the full sequence of packet directions generated by a Tor browser that visits one website at a time, captured from a last-mile network, predict the destination website of the traffic out of a given set of websites''. Note that the task specifies a set of possible traffic objects, packet directions, which is a domain for $X$'s. Similarly, it defines a set of possible websites, which is a range of possible $y$'s. We can imagine a $p(X,y)$ that describes the class probabilities associated with each possible packet direction sequence, if all the Tor browsers in the world were generating traffic towards the specified websites in all the last mile networks in the world. A website fingerprinting dataset, such as the Undefended \cite{dfattack}, which we will visit in this work, is theoretically sampled from $p(X,y)$, but is not necessarily representative of it. 

\textbf{Representativeness}. Dataset representativeness is less straightforward to define. Intuitively, we would call a dataset representative of a task if its input distribution is close to that of the task, \textit{where it matters}, and it is ``correctly'' labeled. A possible definition is that a dataset is representative if the performance of a model evaluated on the dataset is close to the expected value of its performance on $p(X,y)$. However, this definition, ties model generalizability to dataset representativeness, and will lead to non-representativeness and non-generalizability being indistinguishable. We make an attempt at defining representativeness in Appendix \ref{appx:represent-def}. Regardless of the theoretical definition, representativeness will be difficult to verify in practice when $p(X,y)$ is unknown, which includes all practical cases. Hence, in this paper, we compare the distribution of feature values between datasets as a sign of good or bad representativity. For example, if 90\% of the flows collected for a task are shorter than 10 packets and a dataset only contains flows longer than 100 packets, we consider the the dataset non-representative. Note that it is easier to refute representativeness using this method, than to prove it. Moreover, the method assumes the existence of a reference \textit{trusted} dataset. We underline that in our view, representativeness is defined as a relation between a task and a dataset, so a task should be specified before considering dataset representativeness.

\textbf{\textit{Methods} versus \textit{models}.} The definition of generalizability at the beginning of this section defines generalizability for a trained model. However, previously proposed works do not provide a trained model, but describe a recipe, a \textit{method}, with which such a model can be created from a dataset. A method can be applied to any dataset, provided that its required input representation can be extracted from that dataset. For example, website fingerprinting datasets typically only contain time series sequences as traffic data. As a result, traffic classification methods that depend on packet bytes cannot be applied to those datasets. On the other hand, an application identification dataset such as \cite{iscx-vpn} contains packet header bytes and timestamps. A method designed for website fingerprinting can be applied to the dataset, because it is possible to represent the dataset's data as time series sequences and input it to the method. Whether or not the method will perform well on the dataset is, of course, another problem. 

We make this distinction, because for \textit{methods}, generalizability can be defined as their performance remaining within a small threshold of their advertised performance, when they are applied to datasets representing tasks close to the one they were developed to solve. The advantage of method-level generalizability is that it can be studied through same-dataset training and evaluation of the models proposed by those methods. Same-dataset evaluations allow to get around the problem of distribution shift \cite{dl_distributionshift2020}, which is inevitable if a model trained on one dataset is evaluated on another. Distribution shift exists between any two independently collected datasets, since ideal representativeness does not exist in practice, even if two datasets represent the exact same task, which is rarely the case in available traffic datasets anyway. We will get to generalizability for trained models, \ie generalizability in the face of distribution shift, in Section \ref{sec:model-gen-shift}. Before that, however, in Section \ref{sec:dataset-specificity}, we look at method-level generalizability. 

%% file: new_sections/revisiting-prior.tex
\section{Prior methods through the lens of alternative datasets} \label{sec:dataset-specificity}

We revisit three prior methods designed for: 1) Tor website fingerprinting in \cite{dfattack}, 2) application identification on regular, VPN, and Tor traffic in \cite{flowpic}, and 3) service identification on aggregated network traffic in \cite{elham_cnsm}. The methods are called DFattack, Flowpic, and UWTransformer, respectively (See Table \ref{tab:prior-methods}). The methods' underlying tasks resemble each other in that the labels are close, \ie applications, websites and services are closely related in terms of the entities they represent. Moreover, all methods rely on time series features as input representation. These similarities make it a fair comparison.

\renewcommand{\arraystretch}{1.4} 

\begin{table}[h]
\scriptsize
\caption{Properties of revisited methods}\label{tab:prior-methods}
\centering
\begin{threeparttable}
  \begin{tabular}{|>{\raggedright\arraybackslash}p{1.9cm}||
                      >{\raggedright\arraybackslash}p{1.3cm}|
                      >{\raggedright\arraybackslash}p{2.5cm}|
                      >{\raggedright\arraybackslash}p{1.2cm}|}
    \hline
     \textbf{Method} & \textbf{Model Type} & \textbf{Input type} & \textbf{Evaluation Dataset} \\
    \hline
    \hline
    \textbf{DFattack} \cite{dfattack} & 1D-CNN & Packet directions & {Undefended}\\
    \hline
    \textbf{Flowpic} \cite{flowpic, flowpic2023replication} & 2D-CNN (LeNet) & Packet sizes and arrival times & ISCX-Flowpic\\
    \hline
    \textbf{UWTransformer} \cite{elham_cnsm} & Transformer & Packet sizes, directions, and inter-arrival times  & {Orange}
    \\
    \hline
  \end{tabular}
\end{threeparttable}
\end{table}

\textbf{Input Features}. Time series features typically consist of sequences of packets' inter-arrival times (IAT), packet directions and packet sizes, but their representation varies across methods. For example, DFattack takes as input the sequence of packet directions only, cut or zero-padded to a length of 5000. Flowpic builds a pictorial representation of traffic by organizing traffic sequences into snapshots of unidirectional flows, each represented as a grayscale grid. Each cell in the grid shows the number of packets of a specific size and arrival time range. The third method, UWTransformer, takes packet directions, sizes, and IATs, as a 3-channel input, cut or zero-padded to a size of 32. 

\subsection{Datasets and their tasks}

In this section, we evaluate all three methods on one dataset from each work: Undefended from \cite{dfattack}, ISCX-Flowpic selection from \cite{flowpic, iscx-tor}, and Orange (June) from \cite{elham_cnsm}. Table \ref{tab:prior-datasets} summarizes the properties of these datasets. 

\renewcommand{\arraystretch}{1.2} 

\begin{table}[h]
\scriptsize
\caption{Properties of revisited methods' datasets}\label{tab:prior-datasets}
\centering
\begin{threeparttable}
  \begin{tabular}{|>{\raggedright\arraybackslash}p{1.2cm}||
                      >{\raggedright\arraybackslash}p{1.2cm}|
                      >{\raggedright\arraybackslash}p{1.2cm}|
                      >{\raggedright\arraybackslash}p{1.2cm}|
                      >{\raggedright\arraybackslash}p{1.2cm}|}
    \hline
     \textbf{Dataset}  & \textbf{Undefended} \cite{dfattack, tiktok2019} & \textbf{ISCX-Flowpic} \cite{flowpic, miniflowpic} & \textbf{Orange}\tnote{1}  \hspace{1mm} \cite{iman, navid2022datadrift, elham_cnsm} & \textbf{CESNET-25}\tnote{2} \hspace{1mm} \cite{cesnet} \\
    \hline
    \hline
    \textbf{Data unit} & Website trace (bidir) & Flow snapshot (unidir) & Flow (bidir) & Flow (bidir)\\
    \hline
    \textbf{Size} & 105,730 & 7,296 & 51,179 & 1 million\\ 
    \hline
    \textbf{Input type} & Timeseries & Timeseries except direction & Timeseries & Timeseries (First 30) \\
    \hline
    \textbf{Classes} & 95 & 5 & 8 & 25 \\
    \hline
    \textbf{Class type} & Website & Application category & Service category & Service\\
    \hline
    \textbf{Network} & Last mile & Last mile & Aggregation & Aggregation\\
    \hline
    \textbf{Traffic type} &  Tor & Tor &  TLS & TLS\\
    \hline
  \end{tabular}
\begin{tablenotes}
\footnotesize
\item[1] Orange is a series of datasets first introduced in \cite{iman}, but Orange June which is used here only appeared in \cite{navid2022datadrift, elham_cnsm}.
\item[2] CESNET-25 is our selection of \cite{cesnet}'s dataset, which will be used in Section \ref{sec:model-gen-main}.

\end{tablenotes}
\end{threeparttable}
\end{table}

Further dataset descriptions can be found in Appendix \ref{appx:prior-datasets}. Table \ref{tab:prior-datasets} describes the version of each dataset used in this work, while the original dataset might have had multiple types of traffic, labeling, etc.


%% file: new_sections/prior-performance.tex
\subsubsection{Task taxonomy}


For a method to be applied to a new dataset, adjustments are needed to reconcile it to the new task, input representation, and data entity. We found Flowpic's snapshot classification task harder to reconcile with the other two methods' tasks. 
Flowpic takes individual snapshots as input and reports performance at the snapshot level. The underlying task is therefore traffic snapshot classification, as opposed to traffic trace (or flow) classification. We can, of course, break any of the introduced datasets into unidirectional snapshots and run the flowpic method on them. However, the evaluated snapshot-level performances will be incomparable to flow or trace-level. The methods are therefore compared separately for snapshot classification and flow/trace classification.

\subsection{Snapshot classification} \label{sec:snapshot-task}

Snapshot classification is a rather niche task which we only encountered in \cite{flowpic, miniflowpic}, but it provides fair grounds for evaluating Flowpic as its original evaluation task. 
DFattack is excluded from this section's evaluations because it operates on packet directions only, and the snapshots are unidirectional.

\subsubsection{Augmented dataset recreation}. In each dataset, each flow or trace is first broken into at most two unidirectional sequences, which are in turn broken into snippets of equal time durations, as proposed by \cite{flowpic}. Each snippets' time series is then turned into either a picture or an array representation, depending on the method.

The augmented ISCX-Flowpic Tor dataset was recreated from the original dataset shared by \cite{flowpic}, which contains as few as 81 Tor flows. The Tor dataset was nevertheless the subject of multiple evaluations in \cite{flowpic}, and declared the most difficult to classify. Window and step sizes were set to 60 and 15 seconds, as suggested by \cite{flowpic}. Augmentation drastically increases the number of Tor samples because the original flows are long. After augmentation, some classes were undersampled resulting in a class balance very close (but not identical) to the one in \cite{flowpic}. Creating training and test splits without snapshots of the same flow ending up on both sides was nontrivial (\cf Appendix \ref{appx:flowpic-splits}).



The augmentation of Orange was done similarly to ISCX-Flowpic, notably with the same window length and step size, and the same rules regarding not mixing up snapshots with the same origin flow while splitting. However, since the Orange datasets had considerably more flows, a near 5 to 1 train-to-test ratio (85K train snapshots to 21K test) was reached without problem.  

\subsubsection{Method Adjustments.} UWTransformer's number of input channels was reduced from 3 to 2, since packet directions were absent. The input size, 32 in the original work, was increased to 1024, to accommodate the significantly larger packet number in ISCX-Flowpic's snapshots. To give the reader a sense of how different the two datasets are, the median number of packets per snapshot in ISCX-Flowpic is 1733 in the training data, whereas the same number per Orange's unidirectional flow is 16. 

\subsubsection{Results: Flowpic vs UWTransformer} Table \ref{tab:snapshot-flowpic-vs-uwtransformer} shows the accuracy and F1-scores of UWTransformer vs Flowpic, averaged over two experiments. 
The reported F1-score is the average classwise F1-score regardless of class weight. The gap between accuracies and F1-scores measured on ISCX-Flowpic is bigger due to its smaller size combined with its class imbalance. Flowpic achieves an F1-score of 0.09 on chat data, which is supported by 43 snapshots in the test dataset, mistaking 86\% of chat snapshots with browsing, the most frequent label. However, its other classwise F1-scores were all above 0.9. UWTransformer, also has a low average F1-score, with the lowest values 0.25 and 0.38 for video and chat respectively. Incidently, these are also the classes with the fewest training samples among the five. Orange's F1-scores are more uniform, with \textit{games} being the most difficult class for both models with an F1-score of 0.43 and 0.55, in Flowpic and UWTransformer, respectively. Again, this is the class with the fewest number of training samples, 1257 snapshots.

\begin{table}[h]
\scriptsize
\caption{Accuracy for pairs of method and dataset}\label{tab:snapshot-flowpic-vs-uwtransformer}
\centering
\begin{threeparttable}
  \begin{tabular}{|>{\raggedright\arraybackslash}p{1.9cm}||
                      >{\raggedright\arraybackslash}p{1.2cm}|
                      >{\raggedright\arraybackslash}p{1.2cm}|
                      >{\raggedright\arraybackslash}p{1cm}|
                      >{\raggedright\arraybackslash}p{1cm}|}
    \hline
       & \multicolumn{2}{c|}{\textbf{Augmented ISCX-Flowpic}} & \multicolumn{2}{c|}{\textbf{Augmented Orange}} \\
    \cline{2-5}
       & \textbf{Accuracy} & \textbf{F1-score} & \textbf{Accuracy} & \textbf{F1-score} \\
    \hline\hline
    \textbf{Flowpic}  & 0.92\tnote{1} & 0.77\tnote{1} & 0.65 & 0.64 \\
    \hline
    \textbf{UWTransformer} & 0.76 & 0.56 & 0.71 & 0.69 \\
    \hline
  \end{tabular}
  \begin{tablenotes}
    \footnotesize
    \item[1] Slightly different from the numbers reported in \cite{flowpic}, \cf Appendix \ref{appx:flowpic-splits}.
 \end{tablenotes}
\end{threeparttable}
\end{table}

Each model performs better on its original evaluation dataset, as a trend that we will see throughout this work. The gap is smaller on Augmented Orange, possibly because it is not exactly the same evaluation dataset as in \cite{elham_cnsm} and hence model hyperparameters were not chosen based on this dataset. For reference, the reported accuracy of UWTransformer on bidirectional flows in Orange is 0.86. 


\subsubsection{Flowpic's expected performance} Two question were raised by the above results. First, \textit{Which one of the two performances is more representative of Flowpic's performance?} Orange and ISCX-Flowpic are very different datasets in terms of input distribution, and Flowpic's lower performance on Orange may be an outlier. It is possible, for example, that Flowpic performs well on other last-mile script-generated traffic datasets, which Orange does not represent, but ISCX-Flowpic and Undefended do. Hence, evaluating Flowpic on Undefended is key to testing this hypothesis.

Second, even if the Orange evaluation is representative of Flowpic's out-of-the box performance, \textit{is there a quick fix for it, in terms of hyperparameter tuning perhaps?} The question brings us to the choice of \texttt{window\_size}. 

The $\texttt{window\_size}$ value not only directly influences performance by determining dataset size, but also controls distinguishing features of the pictures inputted to the model. A smaller window size allows less information represented in more detail, whereas a larger window size allows more information represented but in summary, \eg line and square shapes in a smaller window size may end up as dots in a larger one. Picture dimensions, set to 32x32 in our experiments in accordance to \cite{miniflowpic, flowpic2023replication}, control the level of summarization.

Table \ref{tab:miscel-flowpic-perf} shows Flowpic's accuracy on augmentations of Undefended with different values of \texttt{window\_size}. The values were chosen based on Undefended's statistics, in which 90\% and 25\% of flows are shorter than 60.3 and 12.4 seconds respectively, approximated with 60 and 10. A smaller window size was also tried for curiosity. The table presents the median number of packets per snapshot as an estimate of the amount of information in each snapshot. Table \ref{tab:miscel-flowpic-perf} suggests that Flowpic's performance is contingent on snapshots containing most of the flow. Even a median packet number of 107 per snapshot is not enough otherwise. 




As shown in Table \ref{tab:miscel-flowpic-perf}, Flowpic's accuracy with the best window size on Undefended is in the 0.60s, and closer to Orange than ISCX-Flowpic. Since most Undefended flows fit within a 60 second snapshot, Flowpic's snapshot-level accuracy can be compared to DFattack's flow-level accuracy on Undefended, a whopping 0.97, with the caveat that Flowpic achieves its accuracy based on only one direction of the flow. Still, the table's results corroborate our hypothesis that proposed methods perform unusually better on their original evaluation datasets, and that those performances do not represent their expected performance.

\renewcommand{\arraystretch}{1.4}  
\begin{table}[t]
\scriptsize
\caption{Preprocessed data properties and performance of Flowpic on Undefended}\label{tab:miscel-flowpic-perf}
\centering
\begin{threeparttable}
  \begin{tabular}{|
                      >{\raggedright\arraybackslash}p{2.1cm}||
                      >{\raggedright\arraybackslash}p{1.4cm}|
                      >{\raggedright\arraybackslash}p{1.6cm}|
                      >{\raggedright\arraybackslash}p{1.6cm}|}
    \hline
      \textbf{Augmented Dataset} & \textbf{Undefended window = 1s}  & \textbf{Undefended window = 10s} & \textbf{Undefended window = 60s}   \\
    \hline
     \textbf{Snapshots} & 4,279,390 & 656,933 & 232,674 \\
    \hline
    \textbf{Median packetnum} & 14 & 107 & 394 \\
    \hline
    \textbf{Accuracy} & $0.12\pm0.005$ & $0.29\pm0.01$ & $0.63\pm 0.01$ \\
    \hline
  \end{tabular}
\end{threeparttable}
\end{table}


%% file: new_sections/prior-performance-contd.tex
\subsection{Flow and trace classification} \label{sec:prev-methods-task2}

\begin{figure}[htb]
  \centering
  \begin{subfigure}{\linewidth}
    \centering
    \includegraphics[width=0.7\linewidth]{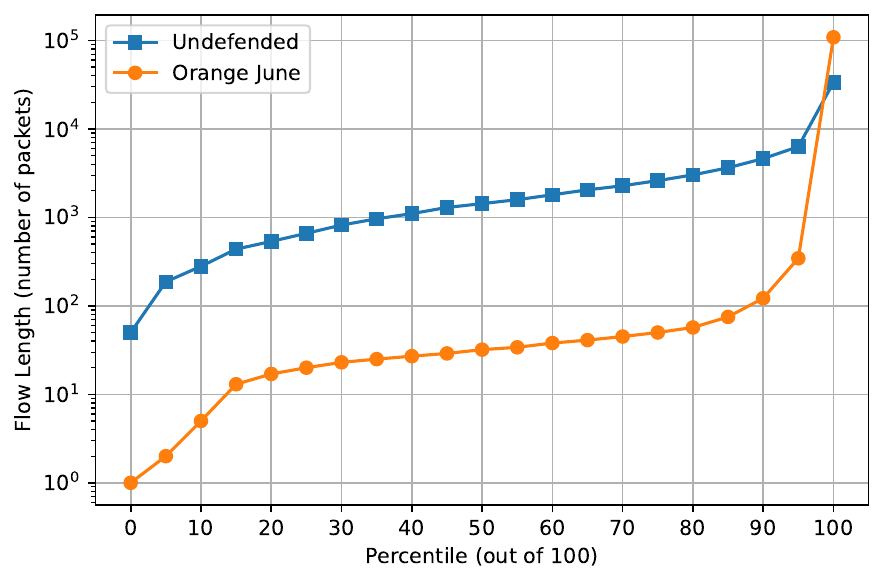}
    \caption{Flow/trace length percentiles for Orange and Undefended}
    \label{fig:ds-percent}
  \end{subfigure}
  
  \vspace{1em} 

  \begin{subfigure}{\linewidth}
    \centering
    \includegraphics[width=0.7\linewidth]{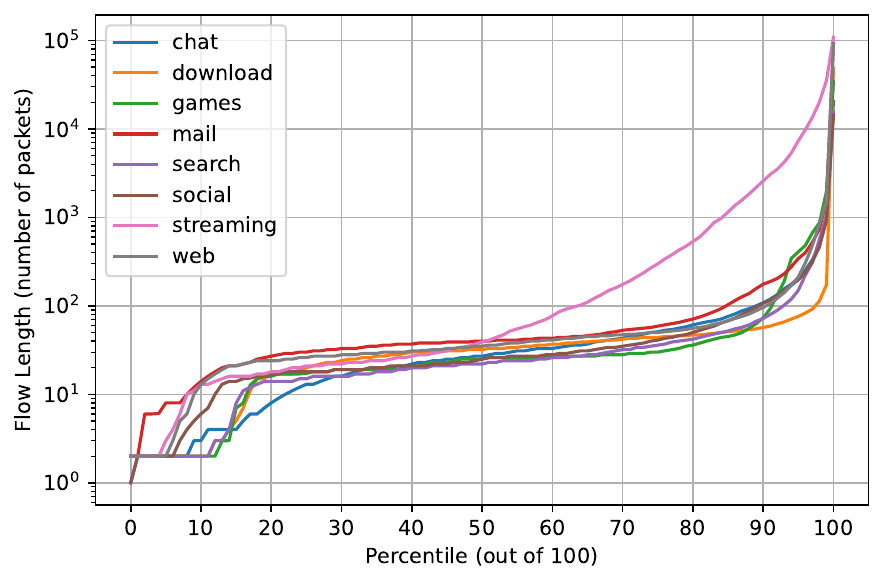}
    \caption{Flow length percentiles for Orange classes}
    \label{fig:orange-cls-percent}
  \end{subfigure}

  \caption{Sequence lengths in Undefended and Orange}
  \label{fig:flow_length_percents}
\end{figure}

The task in this section assumes that the entire flow or trace is available, although the methods may not use all of it. DFattack and UWTransformer both introduce a cutoff point, \ie a maximum number of packets beyond which input sequences are ignored. The cutoff points in the original methods are 5000 and 32 respectively, but were adjusted to 50 for DFattack on Orange and 500 for UWTransformer on Undefended according to each dataset's input characteristics (\cf Figure \ref{fig:flow_length_percents}). The learning rates for each method were also tuned to the dataset. 

The three methods' trace-level accuracy on Undefended and flow-level accuracy on Orange are reported in Table \ref{tab:prior-methods-performance}. For Flowpic, we assume the first downstream snapshot of each flow or trace as its representative snapshot to compute its accuracy (\cf Appendix \ref{app:perf-translation}), which significantly improved its performance compared to its snapshot-level accuracy. \texttt{window\_size} for Undefended was set to 60 seconds. For Orange, 70-second and 240-second snapshots were tried, derived from the 75th and 90th percentiles of Orange flow durations. The results were slightly better for 240 seconds, but within two percentage points of one another in both accuracy and weighted F1-score.  


\begin{table}[h]
\scriptsize
\caption{Accuracy of three methods on Undefended and Orange datasets}\label{tab:prior-methods-performance}
\centering
\begin{threeparttable}
  \begin{tabular}{|>{\raggedright\arraybackslash}p{1.4cm}||
                      >{\raggedright\arraybackslash}p{1.5cm}|
                      >{\raggedright\arraybackslash}p{1.8cm}|
                      >{\raggedright\arraybackslash}p{1.5cm}|}
    \hline
    \textbf{} & \textbf{DFattack} & \textbf{UWTransformer} & \textbf{Flowpic} \\
    \hline
    \hline
    \textbf{Undefended} & $0.98 \pm 0.01$ & $0.81 \pm 0.01$ & $0.63 \pm 0.01$ \\
    \hline
    \textbf{Orange} & $0.71 \pm 0.01$ & $0.86 \pm 0.02$ & $0.67 \pm 0.01$ \\ 
    \hline
  \end{tabular}
\end{threeparttable}
\end{table}




As we can see in Table \ref{tab:prior-methods-performance}, the close-to-perfect accuracy attained by DFattack on its original evaluation dataset does not hold on Orange. Yet, the other two methods' performances have a similar trend, \ie each method's performances on the two datasets are close, but lower on Undefended and higher on Orange. We speculate that the trend may reflect the \emph{difficulty} of each dataset's classification task.


To obtain a clearer estimate of DFattack's expected performance, it seems reasonable to bring in another dataset to evaluate DFattack as we did in the previous section. In fact, we tried to do this using the CESNET dataset (\cf Table \ref{tab:prior-datasets}), which will be introduced in Section \ref{sec:cesnet}. However, neither DFattack nor UWTransformer trained smoothly on it. DFattack simply did not train; both training and validation accuracy remained low, and UWTransformer's validation accuracy remained low while its training accuracy skyrocketed to nearly 1. The problem could not be resolved with static learning rates, or decaying learning rate schedules that start from an effective value. We will leave the explanation of the solution to the next section. Here, we point out that DFattack and UWTransformer show this problem as opposed to Flowpic because they are deeper models, and training deeper models is more difficult.  


\begin{figure}[th]
  \centering
  \includegraphics[width=0.7\linewidth]{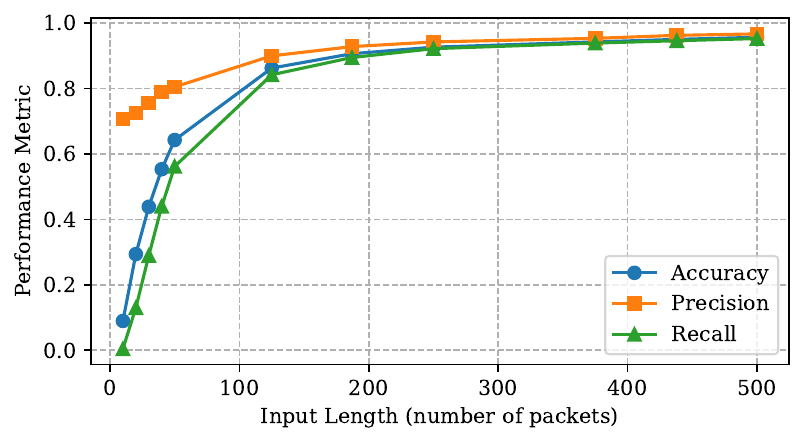}
  \caption{DFattack's performance on Undefended with varying input lengths}
  \label{fig:dfattack-vs-inputlen}
\end{figure}

\textit{Why does DFattack performs so much better on the Undefended compared to Orange?} An important reason is the two datasets' input length distributions. Undefended's traces are considerably longer in terms of number of packets than Orange's flows. 
For example, if Undefended's minimum packet number of 50 was applied to Orange, 75\% of Orange's flows would be left out (\cf Figure \ref{fig:ds-percent}). On the other hand, as Figure \ref{fig:dfattack-vs-inputlen} shows, DFattack's excellent performance on Undefended is only achieved if its input length is over approximately 200 packets, and even on the Undefended dataset, Dfattack's accuracy drops down to 0.64 when its input length is set to 50 as it is on Orange. 

On the other hand, increasing DFattack's input length on Orange (\eg to 500) does not help its performance, because it only lengthens zero paddings of most flows. As Figure \ref{fig:orange-cls-percent} shows, for most classes in the Orange dataset, the above-mentioned input length of 200 is larger than the 90\% percentile. 

\subsection{Discussion}\label{sec:discussion1}
Despite method adjustment practices, \eg finding the best window length, learning rate, input sizes, etc., the results in this section were quite different from what was previously found for each given method. Although obtaining different results for a method on different datasets is expected (after all, some tasks are intrinsically more difficult than others to solve), the observed performance patterns do not readily point to task difficulty as the reason. For example, if one dataset is intrinsically easier than another to classify, we would expect lower performances on it from all methods. Rather, what we saw was each method excelling at its own evaluation dataset with a remarkable margin, as if it were tailored to it.

We saw evidence of seemingly dataset-specific attributes carrying method performances. For example, Flowpic's good performance in classifying ISCX-Flowpic's Tor snapshots did not seem to generalize to other snapshot classification tasks on other Tor datasets. The experiments also showed that large trace lengths, arguably a dataset-specific attribute, is an important factor in Dfattack's performance. On the other hand, methods showed decent performance in solving different \textit{domain-level tasks}, \ie tasks that make sense to the domain expert, such as website fingerprinting or intrusion detection, as all methods achieved above 0.6 accuracy after some adjustments, on all datasets. 

\textit{Do the obtained performance values reflect the best performance for each method and dataset pair?} Probably not. The necessity for some hyperparameter tuning was clear when applying each method to each new dataset, and the searches could be more exhaustive. However, in practice, hyperparameter values are chosen based on datasets different from the ones the method is required to perform on. In the absence of guidelines for finding such values without a ground truth (such as choosing the 90th percentile of flow durations as \texttt{window\_len}), a method's realistic performance will likely be based on subpar hyperparameter values, closer to what we saw in this section than to the best value.


Dataset specificiy in methods is not completely unexpected. After all, method creators make their design choices based on what \textit{works} on the data they \textit{have}. From a different point of view, they suppose the data representative of some task, and search for a model that performs well under the assumption. The question is to what extent the specific attributes of the dataset that make or break performance are relevant to the practical task that it represents. For example, the contingency of DFattack's performance on trace length could be acceptable if Tor traces are typically long. 
In that case, the method can be argued to be task-specific, rather than dataset-specific. Such an argument can only be refuted by experimenting on two datasets representing the exact same task. The next section is an attempt to provide such datasets.

%% file: new_sections/model-generalizability-main.tex
\section{A test framework for traffic model generalizability} \label{sec:model-gen-main}\label{sec:test-framework}


In October 2021, two independent teams collected real-world TLS traffic from two different aggregation networks in different countries, each used by thousands of users. Both traffic datasets were labelled by SNI fields, had bidirectional flows (defined in Appendix \ref{appx:prior-datasets}) as data items, and included time-series features (\eg inter-arrival times, packet directions and sizes), for each flow. The datasets, called CESNET \cite{cesnet} and Orange October \cite{elham_cnsm}, are the core of 
our framework for evaluating trained model generalizability across two independent datasets representing the same task. Such a study was previously conducted \cite{navid2022datadrift, elham_cnsm} on datasets collected from the same network at different points in time. Here, the two datasets are collected from different large-scale networks, but around the same time.

\textit{On test framework philosophy.} Works that address ML model robustness to distribution shift \cite{neyshabur2022unlabeled-ood, arjovsky2019invariant} 
often simulate distribution shift by artificially altering an existing dataset with domain-relevant techniques, similar to those used in data augmentation, \eg rotation and colorizing in vision, sequence shift in timeseries, etc. While this methodology is perfectly valid, it leaves a question as to how similar those synthesized datasets are to distribution shifts encountered in practice. We take a different approach in this work (close to that of \cite{saca2021wilds} in computer vision), as we prioritize finding expected generalizable performance over providing solutions. We align real-world datasets as much as possible based on domain knowledge. The remaining shift is assumed to reflect a realistic distribution shift in traffic data for our task. 

\subsection{Task model}

We imagine a futuristic scenario in which Encrypted SNI and Encrypted Client Hello are widely adopted in TLS traffic. Meanwhile, in some geographical areas, the problem of middleboxes dropping SNI-less TLS connections still persists, leading the same services that offer Encrypted SNI in other areas to still include the SNI in those areas. We call the former the safe area and the latter the unsafe area. The task is to find the destination server name of each flow in the safe area, given training data from the unsafe area. The following are the assumptions: 
\begin{enumerate}
\item The privacy discrepancy between the two areas' TLS protocols does not affect client and server implementations in a way that would affect flow timeseries features, such as TCP payload sizes, packet sequence, etc. However, the feature distributions may still be different due to other factors that change traffic features from one network to another.
\item The classification task's number of classes is 25, \ie the possible server names are already constrained to 25 by IP addresses, or other giveaways. (In practice, this number can be very different depending on the service and its hosting company, \cf \cite{esni2020assessing}.)
\item Abundant labeled traffic is available in the unsafe area for any SNI of interest, but only time-series features are available for each flow. 
\end{enumerate}


\subsection{Dataset alignment} \label{sec:dataset-alignment}

Despite CESNET and Orange's many shared properties, an alignment process was needed to make the inputs as alike as possible, and the labels to reflect the same classes, thus minimizing distribution shift between the two. The alignment can also be viewed as the process with which the Orange datasets were transformed to represent the same \textit{task} as CESNET, by defining the same range of labels and input domain. 

\subsubsection{CESNET}\label{sec:cesnet} A public TLS dataset \cite{cesnet} collected from a network shared among several institutions in the Czech republic with half a million users, CESNET was collected every day for two weeks starting from October 4th, 2021. It includes 142 million flows, each labeled with a tag representing a group of closely related domain names and a broad category, totaling 200 tags and 21 categories including \textit{Other}. It includes statistical and timeseries features for each flow. 

\subsubsection{Orange June and October} Collected in June and October 2021 as part of the Orange datasets (\cf Section \ref{sec:dataset-specificity}), the two datasets contain 320K and 339K flows before labeling, and include header bytes, timeseries and statistical features for each flow. The headers and original SNIs were leveraged to alter the datasets from their original form, which appeared in \cite{iman, elham_cnsm, navid2022datadrift} and Section \ref{sec:dataset-specificity}, and align them to CESNET. 

\subsubsection{Label alignment}

\begin{figure}[h]
    \centering
    \includegraphics[width=0.48\textwidth]{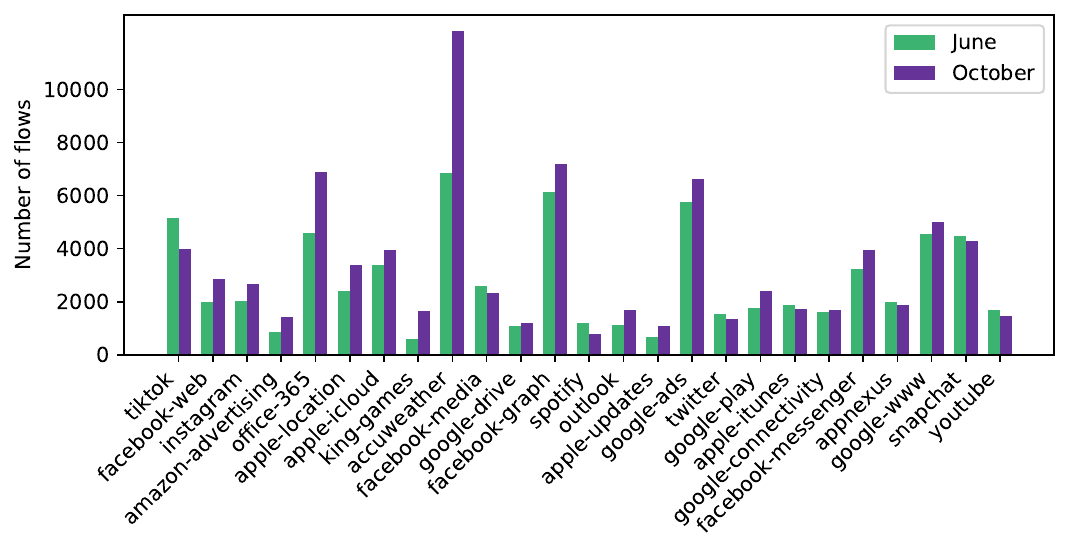} 
    \caption{Label frequencies in Orange datasets}
    \label{fig:orange-balance}
\end{figure}

CESNET and Orange originally had different sets of labels, although their labeling functions both take an SNI value as input and map it to a pair of labels belonging to a two-level label hierarchy. Both also make use of regular expressions in SNI matching. Because the original SNI values were available in Orange but not in CESNET, we applied CESNET's labeling function to Orange's SNIs, which resulted in more than 25\% of the flows labeled under 103 service names, while the rest of CESNET's 200 lower-level labels did not make an appearance. Not all 103 classes had enough flows associated with them to be represented adequately. For example, 46 classes had fewer than 100 flows in either dataset. We selected the 25 classes that happened to be the most frequent in both datasets, 21 of which had more than 1000 samples. Figure \ref{fig:orange-balance} shows the class names and balance of the final Orange dataset selections. 

The same 25 classes were selected from CESNET, but they were undersampled to manage dataset size and to create strictly balanced training and test datasets. Table \ref{tab:num-flows-ds} summarizes the number of flows after each step in the alignment process. CESNET's training-validation and test sets are selected from the first and last three days of collection, respectively, to allow maximum distribution shift. However, evaluated performances were still very close on CESNET's validation and test sets.

\begin{table}[t]
\scriptsize
\caption{Number of flows after each alignment step}\label{tab:num-flows-ds}
\centering
\begin{threeparttable}
  \begin{tabular}{|
                      >{\raggedright\arraybackslash}p{1.3cm}|
                      >{\raggedright\arraybackslash}p{1.1cm}|
                      >{\raggedright\arraybackslash}p{1.3cm}|
                      >{\raggedright\arraybackslash}p{1.8cm}|
                      >{\raggedright\arraybackslash}p{1.1cm}|}
    \hline
      \textbf{Dataset} & \textbf{All Collected} & \textbf{Labellable by CESNET labels}  & \textbf{From 25 most frequent classes in Orange} & \textbf{Selected}   \\
    \hline
    \hline
     \textbf{June} & 320K &  81.8K & 69.1K & 69.1K  \\ 
    \hline
    \textbf{October} &  339K & 97.9K & 83.6K & 83.6K  \\ 
    \hline
    \textbf{CESNET} & Unknown &  141.7 million  & 110 million & 1.2 million\\ 
    \hline
  \end{tabular}
\end{threeparttable}
\end{table}

\subsubsection{Input alignment} 
The original CESNET dataset includes time series sequences only for the first 30 packets of a flow that have a payload. The sequence includes inter-arrival time, direction, and payload size for each non-empty packet. In contrast, Orange timeseries features originally include all packets in a flow, and sizes in the time series reflect packet sizes. Since Orange also includes header bytes for each packet while CESNET doesn't, we extracted CESNET-style time series features for each flow from the Orange dataset.

\subsubsection{Input features, cleaning, and normalization} \label{sec:input-features}
The input alignment's resulting input features are three arrays of inter-arrival times, payload sizes, and packet directions for the first 30 non-empty packets of each bidirectional flow. 

The input IAT and payload size values are normalized by the MinMax function in this section's experiments, unless otherwise stated. Although it is not the best-performing normalization method, MinMax has the advantage of only having a minimum and maximum parameter, which are the same across datasets for payload sizes naturally, and for IATs after we introduced a higher bound. Other normalization methods add complexity to cross-dataset model comparison, as transferred performances may differ significantly depending on which dataset's parameters, \eg mean, standard deviation, etc., are used in the normalization function. 

The distribution of IAT values in CESNET had a very long tail, even in logarithmic scale, which complicated its normalization. As a result, based on the distribution shape, an upper bound of 5 seconds was chosen for both CESNET and Orange, and all larger IAT values were rounded down to 5s. This also took care of the difference in flow timeout times in CESNET and Orange. Also, CESNET had a small number of negative IATs, which we replaced with zero.




\input{new_sections/ml-models}

%% file: new_sections/ml-models.tex
\subsection{Transformer Models} \label{sec:trans-models}



Four transformer models in addition to UWTransformer were created to explore the effect of architectural hyperparameters on model generalizability. They were designed by starting from UWTransformer's architecture and altering architecture choices according to model design conventions. For example, dense layers typically do not come at the beginning of deep models, so the removal of the embedding function, which is a dense layer in UWTransformer, is tested as a choice. Likewise, the number and size of attention heads multiplied together usually equals the input size of the transformer encoder's attention layer, which in the absence of an embedding function in UWTransformer is 3, so the First Model was chosen to have 1 head of size 3. Table \ref{tab:models} summarizes the designed models' differing hyperparameters. The Full Model is an alias for UWTransformer.

\begin{table}[h!]
\scriptsize
\caption{Architectures and hyperparameters}
\centering
\begin{tabular}{|c|c|c|c|c|c|c|}
\hline
\begin{tabular}{@{}c@{}}\textbf{Model} \\ \textbf{Name}\end{tabular} &  \textbf{heads} & \begin{tabular}{@{}c@{}}\textbf{head} \\ \textbf{size}\end{tabular} & \begin{tabular}{@{}c@{}}\textbf{encoder} \\ \textbf{blocks}\end{tabular}  & \begin{tabular}{@{}c@{}}\textbf{Dense} \\ \textbf{units}\end{tabular} &
\begin{tabular}{@{}c@{}} \textbf{embed-}\\ \textbf{ding}\end{tabular} & 
\begin{tabular}{@{}c@{}} \textbf{params}\\ $\times10^3$\end{tabular}\\ 
\hline\hline
First Model & 1 & 3 & 2 & 512 & None & 60\\ 
\hline
More Blocks & 2 & 6 & 4 & 512 & None & 62\\ 
\hline
New Run & 8 & 3 & 2 & 90 & None & 12\\
\hline
NoMB & 8 & 256 & 2 & 512 & None & 122\\
\hline
Full Model & 8 & 256 & 2 & 512 & Dense & 8223\\
\hline
\end{tabular}
\label{tab:models}
\end{table}

The number of parameters (last column in Table \ref{tab:models}) is a result of the choices in the other columns and an input shape of (30,3) shared between all models. It is worth mentioning that the multi-head attention layers in all these transformer models are channel-wise, \ie heads span over and choose between different input dimensions of the same packet rather than multiple packets. Coincidentally, this was suggested as the better choice for timeseries data in \cite{samformer2024}.





%% file: new_sections/gen-experiments.tex
\section{Model generalizability in the presence of distribution shift} \label{sec:model-gen-shift}

Since only two datasets are available for the same task, this work uses the model's \textit{transferred performance} as a measure of its generalizability. Transferred performance is a the performance reached by a model on a \textit{target} dataset, after being trained on a \textit{source} dataset, while source and target datasets are derived from different traffic collections. Same-dataset performance, on the other hand, is evaluated on a separate split of the source dataset, which is not used in training. A smaller difference between transferred and same-dataset performance reflects higher model generalizability. 

In this section, unless otherwise stated, the source dataset is our CESNET selection and the target dataset is either Orange October or Orange June (\cf Table \ref{tab:num-flows-ds}), respectively.

\subsection{Training process} \label{sec:training-process}

The problem of generalizability is only interesting if the model already performs well on the source dataset. However, as mentioned in Section \ref{sec:prev-methods-task2}, training deeper models on CESNET faces obstacles right at the beginning, by showing unstable learning curves that ambiguate model comparison. This is because the same model can demonstrate different performances by different training process choices, notably the choice of optimizer and learning rate schedule.


Transformers are known to be difficult to train \cite{liu2020trainingdifficulty}. Their training curves, \ie loss or accuracy per epoch for the training and validation sets, are closely tied to the choice of optimizer and learning rate schedule \cite{rotational2024equilibrium}, which affects not only convergence speed, but also final converged performance. For example, the New Run model, which at best reaches an accuracy of 0.54 after training with a static learning rate for 140 epochs, can reach an accuracy of 0.73 after only 40 epochs, if trained with a different learning rate schedule, even though both training curves show convergence. Interestingly, we found DFattack, which is not transformer-based, to also require special learning rate schedules to train on CESNET.

\subsubsection{Optimizer} The experiments use Adam \cite{adam}, without any regularization terms, as in UWTransformer. Preliminary experiments tested stochastic gradient descent (SGD) according to \cite{cyclicLR}, because cyclical learning rates perform well for a tree-based model on CESNET according to \cite{cesnet}, and Adamax, DFattack's optimizer. SGD was slower and performed lower than Adam for two tested transformers, and Adamax's performance was slightly lower than Adam. 

\begin{figure}[htb]
  \centering
  \begin{subfigure}{\linewidth}
    \centering
    \includegraphics[width=0.7\linewidth]{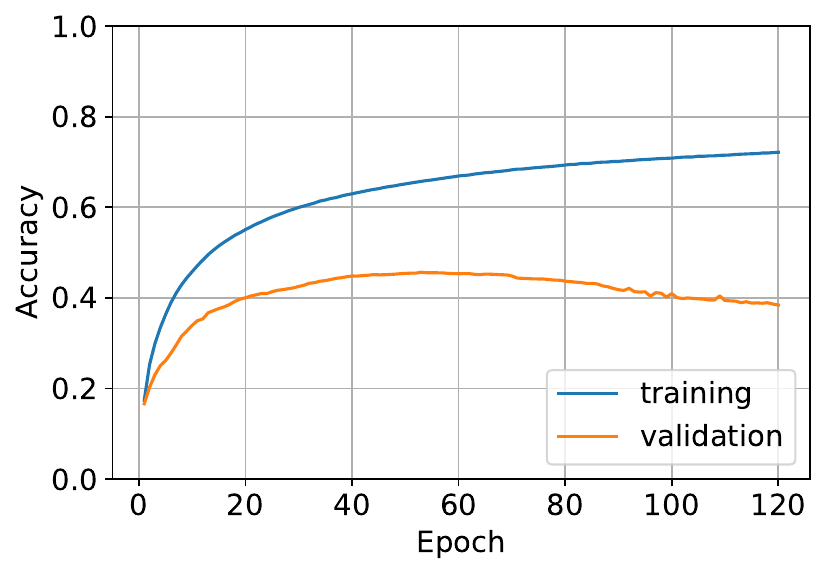}
    \caption{Constant learning rate $=10^{-6}$}
    \label{fig:fm-static}
  \end{subfigure}

  \begin{subfigure}{\linewidth}
    \centering
    \includegraphics[width=0.7\linewidth]{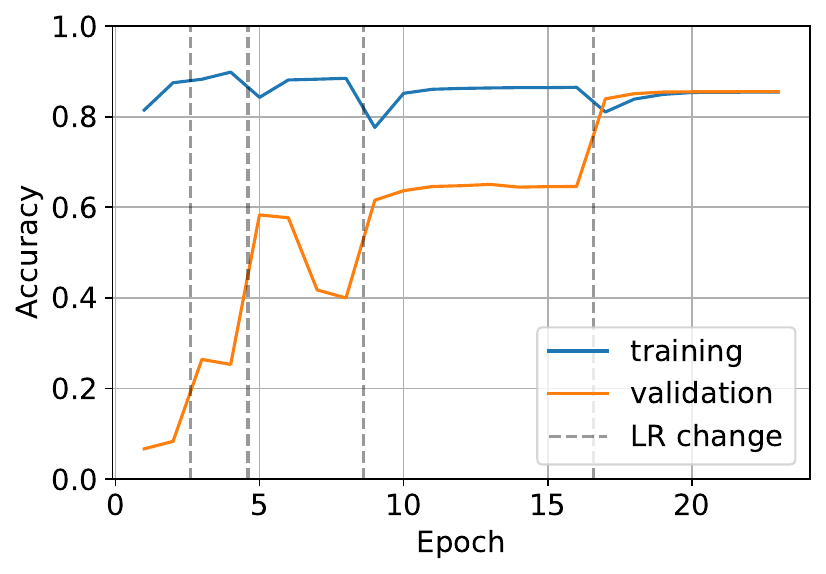}
    \caption{Learning rate series $=[10^{-3},10^{-4},10^{-5},10^{-6}, 10^{-7}]$}
    \label{fig:fm-multi}
  \end{subfigure}

  \caption{Effect of learning rate on First Model's training process}
  \label{fig:first-model-train-proc}
\end{figure}

\subsubsection{Learning rate schedules}\label{sec:lr-schedules} While different learning rate policies have previously been proposed for deep models and specifically transformers, \eg cyclical, decaying, warming up, etc., not all suggestions worked in practice for our data and models. We found that multiple learning rates needed to be involved in the training process for CESNET, and large learning rates, \ie those on the edge of stability, played a crucial role in guiding the subsequent training process. The value of such learning rates depends on the model. Figure \ref{fig:first-model-train-proc} shows the First Model's training curves for two different learning rate schedules. While a constant learning rate leads to a validation accuracy of 0.44 in 120 epochs, a sequence of learning rates spanning over 4 log scales leads to a validation accuracy of over 0.85 in 25 epochs.

The Transformer models did not find any particular gains from learning rate warmup, or cyclical learning rates on CESNET. The best-performing learning rate schedule depends on the dataset, input normalization, and feature selection, as they all affect the loss landscape. For example, a constant learning rate still performed better on the Orange datasets, than any other learning rate schedule we tried.

\input{new_sections/exp-results}

%% file: new_sections/exp-results.tex
\subsection{Results} \label{sec:prior-gen}

Four sets of experimental results are presented in this section. In three out of four, CESNET is the training and Orange October is the target dataset. CESNET serves as the training dataset because it is the larger dataset and strictly balanced, providing 32K training instances from each class. The CESNET training to validation ratio is 4:1, and the model is evaluated on CESNET test set (200K flows) and all of the Orange October selection.

\subsubsection{Previous methods' results} 
\begin{figure}[htb]
  \centering
  \begin{subfigure}{\linewidth}
    \centering
    \includegraphics[width=0.9\linewidth]{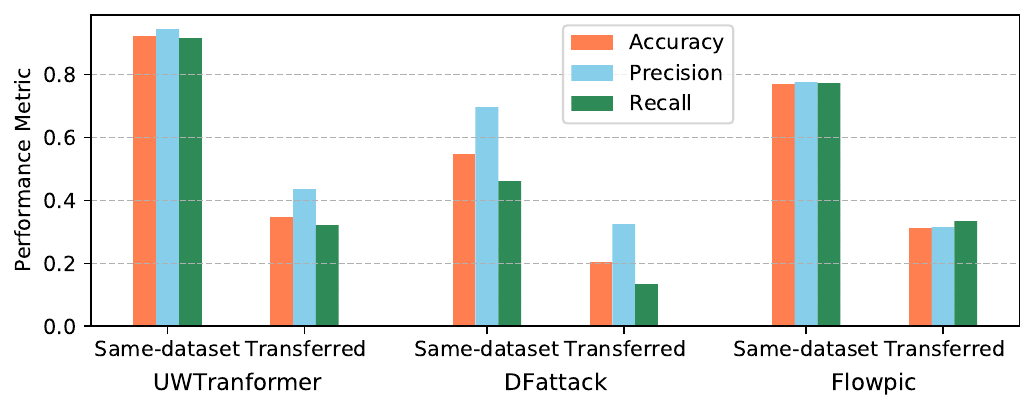}
    \label{fig:transferred-group-task}
  \end{subfigure}

  \caption{Best CESNET to Orange October performance of previous methods}
  \label{fig:gen-prev-methods}
\end{figure}

The first experiment measures trained model generalizability for Section \ref{sec:dataset-specificity}'s previous methods. Figure \ref{fig:gen-prev-methods} shows the same-dataset and transferred performances of the three methods on CESNET and Orange October, respectively. As we can see, despite the two datasets' alignment, a large gap exists between the test performance on CESNET and the transferred performance to Orange, and the best transferred accuracy, UWTransformer's, is 0.35, followed by Flowpic's 0.31.

The results shown in Figure \ref{fig:gen-prev-methods} are indeed the best results seen for each model, and easy to break if different adjustments are made to methods, \eg different normalization for UWTransformer, or a different snapshot-to-flow performance translation for Flowpic. However, the adjustments were made only considering same-dataset performance, to avoid using information that would not be available at training time in practice. 

\textit{Adjustments.} In the case of UWTransformer, we report the performance of the best-performing normalization technique; IAT normalized by a Box-Cox transform fitted on CESNET, packet sizes normalized by MinMax, and packet directions mapped to 0.5 and 1.5. In the case of Flowpic, the first downstream snapshot was used as the flow's representative to turn snapshot-level accuracy to flow-level. \texttt{window\_size} was set to 500 milliseconds (ms), which is between CESNET's 50th and 75th percentiles of flow durations\footnote{Note that the time series durations do not reflect entire flow durations in CESNET.} ($median= 125ms$). Finally, DFattack was trained with Adam as optimizer with learning rates $[10^{-3},10^{-4},10^{-5},10^{-5}]$ each used for the following number of epochs $[3,1,2,10]$. Each model was trained twice under the best adjustments to derive the results.


\textit{Takeaways on feature selection.} Contrary to what \cite{wfrealistictorattack2016} suggests about the benefits of exclusively using packet directions to tackle concept drift in Tor website fingerprinting, Figure \ref{fig:gen-prev-methods}'s results show that this negatively impacts transferred performance, in the context of our task. It also led to lower performance on UWTransformer, whose best measured transferred accuracy on Orange October was 0.20, when only trained on packet directions.

\subsubsection{Effect of model architecture}

\renewcommand{\arraystretch}{1}
\begin{table*}[htb]
\caption{Transformer and classical ML generalizability from CESNET to Orange June and October}
\label{tab:model-perf-trans}
\centering
\begin{tabular}{|c|l|c|c|c|c|c|c|c|c|}
\hline
\textbf{Type} & \textbf{Model} & \multicolumn{2}{|c|}{\textbf{Cesnet Test Set}} & \multicolumn{2}{|c|}{\textbf{Orange June}} & \multicolumn{2}{|c|}{\textbf{Orange October}} & \multicolumn{2}{|c|}{\textbf{Rank}}\\
\cline{3-10}
 &  & \textbf{Accuracy} & \textbf{F1-Score} & \textbf{Accuracy} & \textbf{F1-Score} & \textbf{Accuracy} & \textbf{F1-Score} & \textbf{S} & \textbf{T} \\
\hline\hline
\multirow{5}{*}{Transformer} & First Model      & 0.8500 & 0.8494 & 0.2053 & 0.1996 & 0.2151 & 0.2310 & 4 & 5\\
\cline{2-10}
 & More Blocks    & 0.8101 & 0.8083 & 0.1886 & 0.1953 & 0.1813 & 0.2230 & 6 & 6\\
\cline{2-10}
 & New Run        & 0.7054 & 0.7022 & 0.1814 & 0.1727 & 0.1471 & 0.1646 & 7 & 7\\
\cline{2-10}
 & NoMB      & 0.8441 & 0.8437 & 0.2266 & 0.2255 & 0.2000 & 0.2217 & 5 & 4\\
\cline{2-10}
 & Full Model     & 0.9086 & 0.9274 & \underline{0.3015} & \underline{0.3174} & \underline{0.2767} & \underline{0.3063} &  2 & 1\\
\hline
\multirow{2}{*}{Classical ML} & 1-NN           & 0.8698 & 0.8695 & 0.2590 & 0.2505 & 0.2442 & 0.2633 & 3 & 2\\
\cline{2-10}
 & XGBoost        & \underline{0.9308} & \underline{0.9303} & 0.2235 & 0.2472 & 0.2314 & 0.2628 & 1 & 3\\

 
\hline
\end{tabular}
\end{table*}

The different transformer architectures described in Section \ref{sec:trans-models} were evaluated on the test framework to see if tuning architectural hyperparameters help transferability. The primary hypothesis was that the large number of input dimensions created by the first Dense layer in UWTransformer (\cf Appendix \ref{appx:model-listings}) may be contributing to dataset memorization and the model overfitting to the source dataset. Initial results before our discovering the role of unstable learning rates (\cf Section \ref{sec:training-process})
seemed to support this hypothesis. 
However, after experimenting with learning rate schedules and training models with the schedules that maximized each architecture's same-dataset performance, the evaluated transferred performances showed a different picture. Table \ref{tab:model-perf-trans} shows the performances of the best trained model of each architecture according to CESNET. We see indeed that models with more parameters (\cf Table \ref{tab:models}) tend to have better transferred as well as same-dataset performance.

\begin{figure}[htb]
  \centering
  \includegraphics[width=\linewidth]{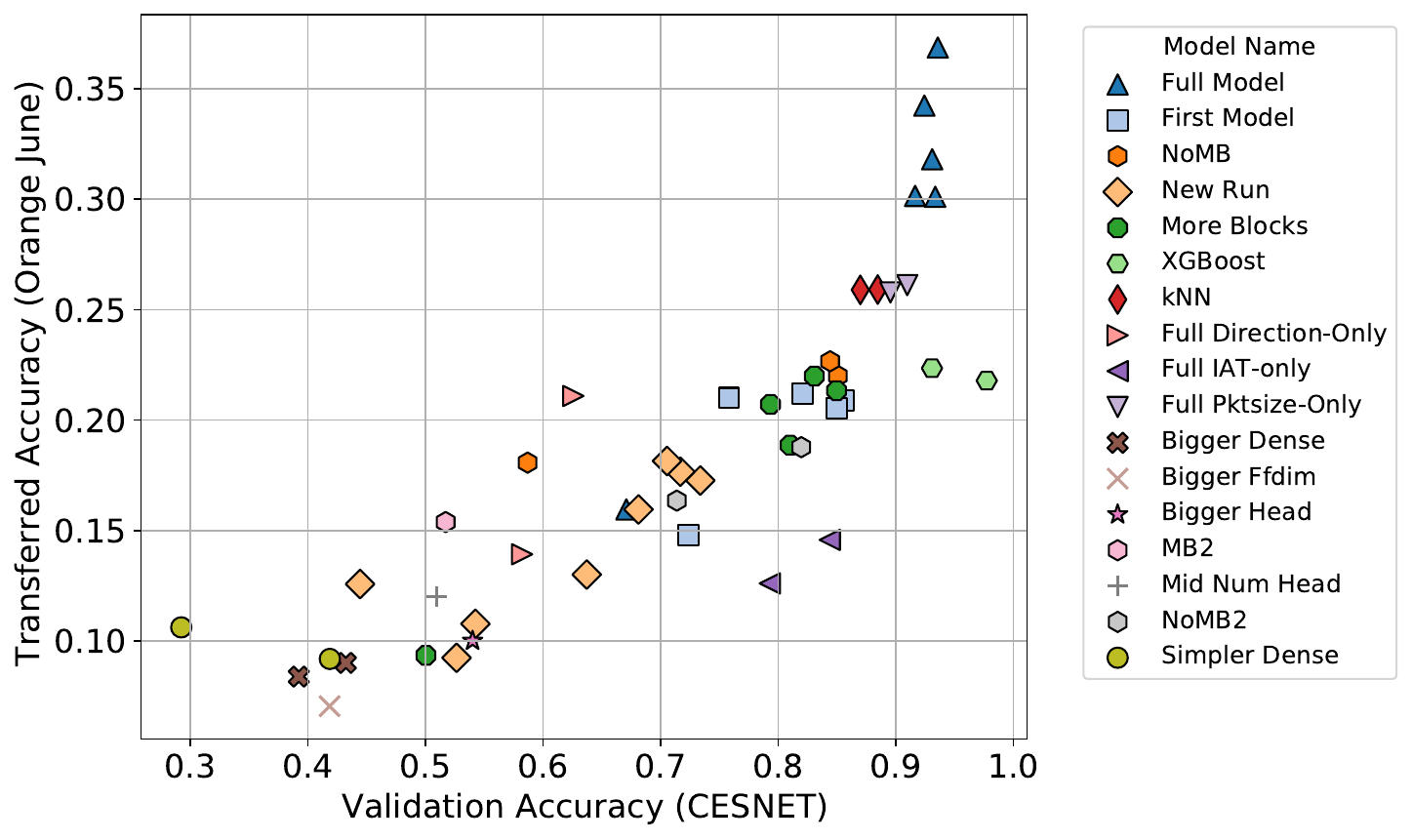}
  \caption{Validation vs. transferred accuracy of different trained models} \label{fig:trained-models-scatter}
  \Description{The same model may show different performances depending on training process. All trained models are shown, not all models' architectures are introduced in the text.}
\end{figure}

Figure \ref{fig:trained-models-scatter} shows transferred vs same-dataset performances for all CESNET-trained models, not just the best. The markers of the same shape and color represent the same architecture trained with different learning rate schedules, and in the case of the Full model, normalization techniques. Some trained models may not have converged, but their training lasted at least 40 epochs. Not all markers are labeled because some represent slightly different architectures from the ones introduced in Section \ref{sec:trans-models}. The figure shows how the same architecture may show different performances with different training processes.

Two classical ML models from the \texttt{scikit-learn} Python library were also trained on CESNET as baseline; an XGBoost model, with the default 100 estimators, and a 1-Nearest Neighbor (1-NN) model, with cosine distance as metric. 

In Table \ref{tab:model-perf-trans}, all models' inputs consists of the same features and MinMax normalization as described in Section \ref{sec:input-features}. We can see that the Full Model, \aka UWTransformer, is the only model that outperforms the classical baselines. 

Another trend that stands out in Table \ref{tab:model-perf-trans} is that even though the gap is large between source and target performances, higher same-dataset performance almost always leads to higher transferred performance, if we ignore differences within 0.01 points. This was an unexpected finding. Although we expected transferred performances to be below same-dataset performance, we did not expect any small accuracy improvement in the source dataset to be reflected in the target\footnote{Note that these are performances after convergence on large datasets representing real traffic. We believe all three conditions matter.}. The last column in the table emphasizes this trend by showing the rank of each model according to its same dataset accuracy (S), and its average accuracy on Orange (T). 

\subsubsection{Effect of task simplification}

\begin{figure}[hb]
  \centering
  \begin{subfigure}{\linewidth}
    \centering
    \includegraphics[width=0.8\linewidth]{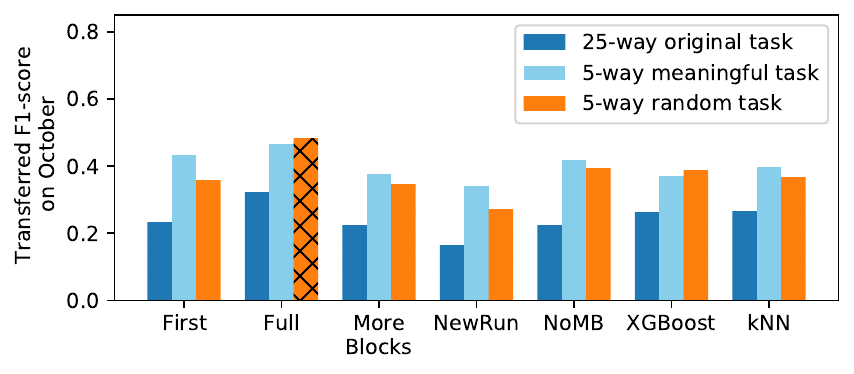}
    \label{fig:transferred-group-task}
  \end{subfigure}
  \begin{subfigure}{\linewidth}
    \centering
    \includegraphics[width=0.8\linewidth]{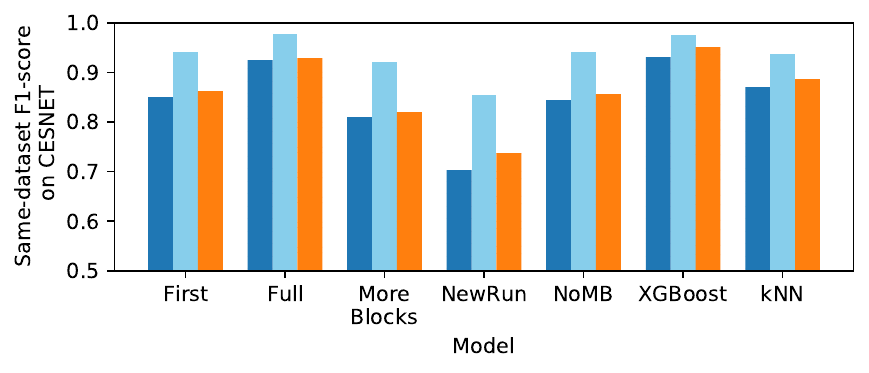}
    \label{fig:cesnet-group-task}
  \end{subfigure}
  \caption{Model performances on tasks with different classes}
  \label{fig:group-task}
\end{figure}

Previous experiments in this section evaluate models on a 25-way classification task, \ie a task involving 25 classes. An examination of confusion matrices in the 25-way task suggested that classes representing similar SNIs tend to be mistaken for one another. For example, \texttt{google-ads} was frequently mistaken for \texttt{google-play} and \texttt{google-www}. Hence, the third set of experiments examines the effect of task label grouping on model performance.

A new \textit{meaningful task} was created, in which each class aggregates inter-related classes from the 25-way task (\cf Figure \ref{fig:orange-balance}) as follows:
\begin{enumerate}
    \item \textit{Facebook}: Labels starting with \textit{facebook}, and instagram
    \item \textit{Google}:  Labels starting with \textit{google} and youtube
    \item \textit{Microsoft}: office-365 and outlook 
    \item \textit{Apple}: Labels starting with \textit{apple}
    \item \textit{Other}: the remaining classes
\end{enumerate}

As a control sample, a different task was created in which labels were grouped randomly, such that the final classes roughly have the same balance as the meaningful task.

Figure \ref{fig:group-task} shows the performances of Table \ref{tab:model-perf-trans}'s models, when evaluated on the two tasks. The original 25-class performance is included for comparison. The models show better performance at identifying groups of related labels, which confirms our hypothesis. The two tasks' difference is, however, less pronounced in transferred performances. Interestingly, the better-performing models are the only two that find the random task easier than the meaningful task on the target dataset. 

\subsubsection{Different training set}
To confirm the transferred performance value ranges, we conducted transfer experiments in the opposite sense, by training on Orange October and evaluating on CESNET test set, for the three best performing models in Table \ref{tab:model-perf-trans}. The transferred accruacies were up to 0.05 better for UWTransformer and XGBoost, while kNN's accuracy remained at 0.25 (\cf Table \ref{tab:orange-to-cesnet} in Appendix \ref{appx:orange-to-cesnet}).




%% file: new_sections/conclusion.tex
\section{Conclusion}\label{sec:conclusion}

Contrary to a common trend in the literature in which the proposed method is the main focus and datasets are there insofar as demonstrating the method's capabilities, in this work, we took a different approach. We looked at datasets as representative of a real-world task, and at classifiers as solutions for the problem posed by the task. We found tuning the solutions to the problem not straightforward, \eg tuning some parameters required knowing the labels (\ie the solution) in advance. We also found that traffic selection criteria, such as cutoff points, need further attention in the literature, as they greatly affect both representativeness and method performance.

Our test framework, whose development was made possible by the coincidental simultaneous collection of real-world traffic by two independent teams, demonstrated a shortcoming in traffic classification which was previously only speculated on; that concept drift is not the only reason for performance gaps between different datasets, that high performances are contingent on a nonrepresentative homogeneity of training and evaluation datasets, and that the task-related part of the knowledge learned by models --rather than the dataset-specific part-- is not enough for reliable use. 

Our evaluations underlined how the same model can reach very different performances on the same dataset, depending on training process and input representation choices. We could see that these differences can exceed the differences between different models by comparing all measured performances with the best measured performances shown respectively in Figure \ref{fig:trained-models-scatter} and Table \ref{tab:model-perf-trans}.



This work's objective was to bring researchers' view of traffic classification closer to the practical view, by differentiating between average vs best-case performances, and showing instances of uncertain test conditions that make reaching best-case performances unlikely, and underlines the need for assessing models based on their average-case performance. We hope to have contributed to the way in which future classifiers are evaluated.

%% file: new_sections/appendix1.tex
\section{Definitions} \label{appx:represent-def}
\textbf{Definition of representativeness}. Consider the space of one-to-one mappings from $part(X)$ to the $x$'s in dataset $D$, where $part(X)$ is the set of all possible partitionings of $X \in p(X,y)$, known as the domain of $x$'s in $p(X,y)$. A representation function is defined as a surjective function that maps each partition in $Part_i(X) \in Part(X)$ to a $x \in D$, and covers $Part_i(X)$. We say $x_i \in D$ represents $X_i \in Part_i(X)$ under representation function $f$ if $f$ maps $X_i$ to $x_i$. We define $D$ as representative of $p$ if there exists a $Part_i(X)$ and a representation function $f$, for which an algorithm obtains perfect accuracy on the task specified by $p(X,y)$ if it predicts $y_i$ as the label of every $x \in Part_i(X)$ for each $(x_i,y_i) \in D$ if $(Part_i(X), x_i) \in f$. Since $p(X,y)$ returns a probability distribution over all $y$'s and not a single $y_i$, the definition assumes maximum likelihood, \ie perfect accuracy over the task means the most likely $y$ for each $x$ is the correct label. Also, it is worth mentioning that by this definition, representativeness is more difficult to realize for some distributions than others, and it may not be possible in some cases. The ease of finding such partitioning depends on the theoretical \emph{complexity} of a task.

\section{Prior works' datasets} \label{appx:prior-datasets}

\subsection{Undefended} The dataset was created by a simulator script visiting each of 95 popular websites' first pages from different devices using a Tor browser. Each website was visited more than 1000 times, resulting in 105,730 traffic sequences in the dataset. Even though DFattack only relies on directions, each traffic sequence in the dataset consists of the full time series features, represented as two stacked arrays; packet arrival times, and signed packet sizes, in which the sign reflects the direction. 
Each data item reflects the traffic sequence generated by visiting a website,\ie a Website trace, which is different from traffic \textit{flows} that we will see in the other two datasets. In effect, each website trace could consist of multiple interwoven flows between the same or multiple IP addresses, all referenced by the target website.  


\subsection{ISCX-Flowpic} The dataset was introduced by \cite{flowpic}, and subsequently used by \cite{miniflowpic}, however, it was left out of a replication of the former in \cite{flowpic2023replication}. It consists of a selection of two previous widely-used datasets \cite{iscx-tor, iscx-vpn} to which \citet{flowpic} added some of their own collected \textit{Chat} traffic. The base datasets \cite{iscx-tor, iscx-vpn} contain VPN, Tor, and regular traffic generated by scripts and collected from a client-side network. \cite{flowpic}'s selection filters out flows with less than 100 packets \cite{miniflowpic}. The dataset selection features 5 application categories and three types of traffic, which amounts to 14 classes in total (because VPN traffic was missing for the \textit{Browsing} class). The paper \cite{flowpic} presents results for different tasks, out of which we chose Tor traffic categorization (5 classes), because the task's theme is close to service identification, and because it was the most challenging task for Flowpic according to the authors. 

Since the size of the original dataset is small (it contains only 183 flows for 12 classes aside from two regular classes), \cite{flowpic} proposes a data augmentation method to increase the size of the small original dataset by moving a window over each flow and taking snapshots of the flow. All the snapshots have the same duration, \eg 60 seconds, and may overlap depending on a step size, \eg 15 seconds. The resulting snapshots are then over-sampled and under-sampled to create a nearly-balanced augmented dataset. 

The data items in the original ISCX-Flowpic selection and the augmented dataset represent different entities. In the original dataset, each data item represents a unidirectional flow. A unidirectional flow is a sequence of packets that share the following five-tuple: Source IP, Destination IP, Source port, Destination port, Protocol, which travel close to each other in time. In the augmented dataset, each data item reflects a fixed-length snapshot of a unidirectional flow, where length refers to a duration of time. As it can be inferred, translating snapshot-level performance to flow-level performance, which is essential in comparisons, is not trivial. The authors implicitly refer to this when comparing their method to others, but make no attempt at providing flow-level performance, possibly because the original dataset is both small and massively imbalanced, so any flow-level performance metric is unlikely to be meaningful. The small size and imbalance can also raise serious questions about its representativeness, however, we will not get into it in this paper, as the quality of the base CIC datasets from which the selection was derived is already put into question \cite{trustee, iman}. We only include the dataset to examine its properties that have led to high measured performances on it.



\subsection{Orange} \label{sec:old-orange-desc}
The Orange datasets are a series of datasets introduced in \cite{iman} and revisited by \cite{navid2022datadrift, elham_cnsm}. They were collected from the mobile network of Orange, an ISP in Europe. The June dataset was derived from an hour-long packet capture in 2021. The collection mechanism included a filter on the client devices to manage the trace size, such that all or none of the traffic of each client device in the time frame appears in the trace. The packet trace was broken into flows using YAF \cite{yaf} according to \cite{iman}, which resulted in 320K flows in total. The dataset is labeled by SNIs, which can be mapped to 8 service classes using a method described by \cite{iman}, resulting in 51K labeled flows. Each data item in the dataset is a bidirectional flow, which is the same as the five-tuple unidirectional flows described before, except packets' source and destination (IP,port) tuple may be switched. The packets in a flow can be at most \textit{flow\_idle\_timeout} seconds apart, which was set to 600 seconds in YAF.

\section{Notes on Flowpic recreation}\label{appx:flowpic-splits}
The 9 to 1 training to test ratio mentioned in \cite{flowpic} seems unreasonable for the selection's Tor dataset, as snapshots derived from the same flow should not end up on different sides of the split, and, some classes contain very few (\eg as few as 6) flows so that one flow ending up on a different side of the split dramatically changes the ratio of snapshots. By trying different random seeds and paying attention to two classes with only 6 and 7 flows, we ended up with a train-to-test ratio of 5279 to 2017 snapshots. 

We were able to achieve results close enough to those presented in the paper for Tor traffic categorization, using this split. We did not attempt cross-validation, as we only use the dataset for model comparison, and the difference between the models is dramatic enough to transcend alterations made by cross-validation. It is worth mentioning that the test set is separate from the validation set. The latter is derived by splitting the training data with a ratio of 9 to 1, with no regard to origin flows. 

We note that the performance measurements reported in Table \ref{tab:snapshot-flowpic-vs-uwtransformer} for Flowpic are slightly different from the ones reported in \cite{flowpic}, possibly for two reasons. While \cite{flowpic} states the window length and step size to be 60 and 15 seconds at the beginning of the paper, for the specific experiments which involve 5-class traffic categorization as above, the window length is set to 15 seconds. Another reason is a difference in  splits, described above. The number ranges they report, $0.87 \pm 0.07$ and $0.65 \pm 0.11$ for accuracy and F1-score respectively, contain the accuracy and F1-score reported above. Similarly to our observations, they show a small F1-score for one of the classes in the confusion matrix.

\section{Snapshot-level to flow-level performance translation} \label{app:perf-translation}

Specifically for Flowpic, we need a way to turn snapshot-level performance to flow-level. The question comes down to when to consider a flow ``correctly'' classified when only predictions for snapshots are available, and a flow is associated with multiple snapshots, each of which may have a different predicted label. Different metrics (described below) were evaluated choosing the last metric in the list. Each list item is a possible choice for $X$ in the following statement: A flow is correctly classified if and only if $X$ of its snapshots are correctly classified, where $X$ equals:
\begin{enumerate}
    \item all 
    \item all downstream
    \item all upstream
    \item most
    \item First two, one in each direction,
    \item First one, preferably downstream,
\end{enumerate}

The numbers issued by different metrics are very different as one may expect. As an example, for a Flowpic model trained on Orange with 70-second snapshots, the accuracies based on the metrics in the order mentioned above are 0.3947, 0.5658, 0.4965, 0.4605, 0.4845, 0.6684. The numbers are reached after 60 epochs of training. We consider it converged at epoch 40 because the last 20 epochs added less than 0.02 accuracy to snapshot-level convergence. Snapshot-level accuracy and macro F1-score for the same model are 0.5923 and 0.5673. 

As we can see in the example, the chosen metric, ``First (preferably downstream) snapshot'' yields the best performance and has the advantage of making sense to a domain expert. It also offers the advantage of associating a flow with a single prediction, which allows the computation of precision and recall performance metrics. In all considered datasets, ``First (preferably downstream)'' yields the best Flowpic performance among all options, except for Undefended, for which ``all upstream'' outperforms it by 0.03 (0.6623 vs 0.6315 in one Flowpic model's validation accuracies). We report the same metric for all datasets in section \ref{sec:prev-methods-task2}, however, because we believe downstream snapshots are a better indicator of a website's behavior than upstream ones, from a domain expert's point of view.



\section{UWTransformer vs DFattack's model structure} \label{appx:why-trans}

From a model structure point of view, UWTransformer has a series of 1-dimensional Convolutional network (1D-CNN) layers, with attention layers in between them that roughly serve as feature combiners and selectors within the scope of every packet's features. The attention layers are short-circuited by residual connections, which gives the model the capacity to ignore their output if necessary. Hence, we view the explored Transformer-based architectures as an upgrade to the well-established 1D-CNNs used in different forms in traffic classification and website fingerprinting \cite{aceto2019DL, iman, aceto2019mimetic, wftriplet}. Compared to DFattack, UWTransformer also has the advantage of using multiple features, which Section \ref{sec:prior-gen} shows works in favor of cross-dataset performance. In terms of computation times, transformers are more costly than 1D-CNNs, but still reasonably fast on a GPU. For example, the average computation time for training on a batch size of 64 is 18 milliseconds for UWTransformer and 5 milliseconds for DFattack, on our GPU.

\section{Orange to CESNET experiment results} \label{appx:orange-to-cesnet}

\begin{table}[htb]
\caption{Generalizability from Orange to CESNET}
\label{tab:orange-to-cesnet}
\centering
\begin{threeparttable}
\begin{tabular}{|c|c|c|c|c|c|c|}
\hline
\textbf{Model} & \multicolumn{3}{|c|}{\textbf{Orange October}} &  \multicolumn{3}{|c|}{\textbf{CESNET Test}} \\ 
\cline{2-7}
 &  Acc. & Prec. & Rec. & Acc. &Prec. &Rec. \\
\hline
\hline
UWTran    & 0.875 & 0.875 & 0.856  & 0.406 & 0.414 & 0.401 \\
\hline
kNN    &   0.785 & 0.775 &  0.750 &   0.252 & 0.268 &  0.252  \\

\hline
XGBoost    & 0.905   &  0.918   &  0.888   & 0.288 & 0.363 &0.288\\

\hline
\end{tabular}
\end{threeparttable}
\end{table}

%% file: new_sections/appendix-model-listing.tex
\section{Model Listings} \label{appx:model-listings}

\begin{table}[h]
\caption{The First Model}
\centering
\small
\renewcommand{\arraystretch}{1.2}
\begin{tabular}{@{}lll@{}}
\toprule
\textbf{Layer} & \textbf{Output Shape} & \textbf{Parameters} \\
\midrule
InputLayer-1 & (None, 30, 3) & 0 \\
AttentionModelTrunk-2 & (None, 90) & 1,022 \\
Flatten-3 & (None, 90) & 0 \\
Dropout-4 & (None, 90) & 0 \\
Dense-5 & (None, 512) & 46,592 \\
Dropout-6 & (None, 512) & 0 \\
Dense-7 & (None, 25) & 12,825 \\
\midrule
\textbf{Total params}: & 60,439 & \\
\textbf{Trainable params}: & 60,439 & \\
\textbf{Non-trainable params}: & 0 & \\
\bottomrule
\end{tabular}
\end{table}

\begin{table}[H]
\caption{Attention Model Trunk of First Model}
\centering
\small
\renewcommand{\arraystretch}{1.2}
\begin{tabular}{@{}lll@{}}
\toprule
\textbf{Layer} & \textbf{Output Shape} & \textbf{Parameters} \\
\midrule
MultiHeadAttention-1 & (None, 30, 3) & 48 \\
Dropout-2 &  & 0 \\
LayerNormalization-3 & (None, 30, 3) & 6 \\
Conv1D-4 & (None, 30, 64) & 256 \\
Dropout-5 &  & 0 \\
LayerNormalization-6 & (None, 30, 3) & 6 \\
Conv1D-7 & (None, 30, 3) & 195 \\
\midrule
\textbf{Total params}: & 511 & \\
\textbf{Trainable params}: & 511 & \\
\textbf{Non-trainable params}: & 0 & \\
\bottomrule
\end{tabular}
\end{table}

\begin{table}[H]
\caption{The MoreBlocks Model}
\centering
\small
\renewcommand{\arraystretch}{1.2}
\begin{tabular}{@{}lll@{}}
\toprule
\textbf{Layer} & \textbf{Output Shape} & \textbf{Parameters} \\
\midrule
InputLayer-1 & (None, 30, 3) & 0 \\
AttentionModelTrunk-2 & (None, 90) & 2,584 \\
Dropout-3 & (None, 90) & 0 \\
Dense-4 & (None, 512) & 46,592 \\
Dropout-5 & (None, 512) & 0 \\
Dense-6 & (None, 25) & 12,825 \\
\midrule
\textbf{Total params}: & 62,001 & \\
\textbf{Trainable params}: & 62,001 & \\
\textbf{Non-trainable params}: & 0 & \\
\bottomrule
\end{tabular}
\end{table}

\begin{table}[H]
\caption{Attention Model Trunk of MoreBlocks Model}
\centering
\small
\renewcommand{\arraystretch}{1.2}
\begin{tabular}{@{}lll@{}}
\toprule
\textbf{Layer} & \textbf{Output Shape} & \textbf{Parameters} \\
\midrule
MultiHeadAttention-1 & (None, 30, 3) & 183 \\
Dropout-2 &  & 0 \\
LayerNormalization-3 & (None, 30, 3) & 6 \\
Conv1D-4 & (None, 30, 64) & 256 \\
Dropout-5 &  & 0 \\
LayerNormalization-6 & (None, 30, 3) & 6 \\
Conv1D-7 & (None, 30, 3) & 195 \\
\midrule
\textbf{Total params}: & 646 & \\
\textbf{Trainable params}: & 646 & \\
\textbf{Non-trainable params}: & 0 & \\
\bottomrule
\end{tabular}
\end{table}

\begin{table}[H]
\caption{The NoMB Model}
\centering
\small
\renewcommand{\arraystretch}{1.2}
\begin{tabular}{@{}lll@{}}
\toprule
\textbf{Layer} & \textbf{Output Shape} & \textbf{Parameters} \\
\midrule
InputLayer-1 & (None, 30, 3) & 0 \\
BatchNormalization-2 & (None, 30, 3) & 12 \\
MultiHeadAttention-3 & (None, 30, 3) & 30,723 \\
Dropout-4 & (None, 30, 3) & 0 \\
Add-5 & (None, 30, 3) & 0 \\
LayerNormalization-6 & (None, 30, 3) & 6 \\
Conv1D-7 & (None, 30, 64) & 256 \\
Dropout-8 & (None, 30, 64) & 0 \\
Conv1D-9 & (None, 30, 3) & 195 \\
Add-10 & (None, 30, 3) & 0 \\
LayerNormalization-11 & (None, 30, 3) & 6 \\
MultiHeadAttention-12 & (None, 30, 3) & 30,723 \\
Dropout-13 & (None, 30, 3) & 0 \\
Add-14 & (None, 30, 3) & 0 \\
LayerNormalization-15 & (None, 30, 3) & 6 \\
Conv1D-16 & (None, 30, 64) & 256 \\
Dropout-17 & (None, 30, 64) & 0 \\
Conv1D-18 & (None, 30, 3) & 195 \\
Add-19 & (None, 30, 3) & 0 \\
LayerNormalization-20 & (None, 30, 3) & 6 \\
Flatten-21 & (None, 90) & 0 \\
Dense-22 & (None, 512) & 46,592 \\
Dropout-23 & (None, 512) & 0 \\
Dense-24 & (None, 25) & 12,825 \\
\midrule
\textbf{Total params}: & 121,801 & \\
\textbf{Trainable params}: & 121,795 & \\
\textbf{Non-trainable params}: & 6 & \\
\bottomrule
\end{tabular}
\end{table}

\begin{table}[H]
\caption{The NewRun Model}
\centering
\small
\renewcommand{\arraystretch}{1.2}
\begin{tabular}{@{}lll@{}}
\toprule
\textbf{Layer} & \textbf{Output Shape} & \textbf{Parameters} \\
\midrule
InputLayer-1 & (None, 30, 3) & 0 \\
MultiHeadAttention-2 & (None, 30, 3) & 363 \\
Dropout-3 & (None, 30, 3) & 0 \\
Add-4 & (None, 30, 3) & 0 \\
LayerNormalization-5 & (None, 30, 3) & 6 \\
Conv1D-6 & (None, 30, 64) & 256 \\
Dropout-7 & (None, 30, 64) & 0 \\
Conv1D-8 & (None, 30, 3) & 195 \\
Add-9 & (None, 30, 3) & 0 \\
LayerNormalization-10 & (None, 30, 3) & 6 \\
MultiHeadAttention-11 & (None, 30, 3) & 363 \\
Dropout-12 & (None, 30, 3) & 0 \\
Add-13 & (None, 30, 3) & 0 \\
LayerNormalization-14 & (None, 30, 3) & 6 \\
Conv1D-15 & (None, 30, 64) & 256 \\
Dropout-16 & (None, 30, 64) & 0 \\
Conv1D-17 & (None, 30, 3) & 195 \\
Add-18 & (None, 30, 3) & 0 \\
LayerNormalization-19 & (None, 30, 3) & 6 \\
Flatten-20 & (None, 90) & 0 \\
Dropout-21 & (None, 90) & 0 \\
Dense-22 & (None, 90) & 8,190 \\
Dropout-23 & (None, 90) & 0 \\
Dense-24 & (None, 25) & 2,275 \\
\midrule
\textbf{Total params}: & 12,117 & \\
\textbf{Trainable params}: & 12,117 & \\
\textbf{Non-trainable params}: & 0 & \\
\bottomrule
\end{tabular}
\end{table}

\begin{table}[H]
\caption{The Full Model (UWTransformer)}
\centering
\small
\renewcommand{\arraystretch}{1.2}
\begin{tabular}{@{}lll@{}}
\toprule
\textbf{Layer} & \textbf{Output Shape} & \textbf{Parameters} \\
\midrule
InputLayer-1 & (None, 30, 3) & 0 \\
Dense-2 & (None, 30, 256) & 1,024 \\
BatchNormalization-3 & (None, 30, 256) & 1,024 \\
Add-4 & (None, 30, 256) & 0 \\
MultiHeadAttention-5 & (None, 30, 256) & 2,103,552 \\
Dropout-6 & (None, 30, 256) & 0 \\
Add-7 & (None, 30, 256) & 0 \\
LayerNormalization-8 & (None, 30, 256) & 512 \\
Conv1D-9 & (None, 30, 64) & 16,448 \\
Dropout-10 & (None, 30, 64) & 0 \\
Conv1D-11 & (None, 30, 256) & 16,640 \\
Add-12 & (None, 30, 256) & 0 \\
LayerNormalization-13 & (None, 30, 256) & 512 \\
MultiHeadAttention-14 & (None, 30, 256) & 2,103,552 \\
Dropout-15 & (None, 30, 256) & 0 \\
Add-16 & (None, 30, 256) & 0 \\
LayerNormalization-17 & (None, 30, 256) & 512 \\
Conv1D-18 & (None, 30, 64) & 16,448 \\
Dropout-19 & (None, 30, 64) & 0 \\
Conv1D-20 & (None, 30, 256) & 16,640 \\
Add-21 & (None, 30, 256) & 0 \\
LayerNormalization-22 & (None, 30, 256) & 512 \\
Flatten-23 & (None, 7680) & 0 \\
Dense-24 & (None, 512) & 3,932,672 \\
Dropout-25 & (None, 512) & 0 \\
Dense-26 & (None, 25) & 12,825 \\
\midrule
\textbf{Total params}: & 8,222,873 & \\
\textbf{Trainable params}: & 8,222,361 & \\
\textbf{Non-trainable params}: & 512 & \\
\bottomrule
\end{tabular}
\end{table}

%% file: onetask_main.bbl

\begin{thebibliography}{58}


\ifx \showCODEN    \undefined \def \showCODEN     #1{\unskip}     \fi
\ifx \showISBNx    \undefined \def \showISBNx     #1{\unskip}     \fi
\ifx \showISBNxiii \undefined \def \showISBNxiii  #1{\unskip}     \fi
\ifx \showISSN     \undefined \def \showISSN      #1{\unskip}     \fi
\ifx \showLCCN     \undefined \def \showLCCN      #1{\unskip}     \fi
\ifx \shownote     \undefined \def \shownote      #1{#1}          \fi
\ifx \showarticletitle \undefined \def \showarticletitle #1{#1}   \fi
\ifx \showURL      \undefined \def \showURL       {\relax}        \fi
\providecommand\bibfield[2]{#2}
\providecommand\bibinfo[2]{#2}
\providecommand\natexlab[1]{#1}
\providecommand\showeprint[2][]{arXiv:#2}

\bibitem[Aceto et~al\mbox{.}(2019a)]%
        {aceto2019DL}
\bibfield{author}{\bibinfo{person}{Giuseppe Aceto}, \bibinfo{person}{Domenico Ciuonzo}, {et~al\mbox{.}}} \bibinfo{year}{2019}\natexlab{a}.
\newblock \showarticletitle{Mobile encrypted traffic classification using deep learning: Experimental evaluation, lessons learned, and challenges}.
\newblock \bibinfo{journal}{\emph{IEEE Transactions on Network and Service Management}} \bibinfo{volume}{16}, \bibinfo{number}{2} (\bibinfo{year}{2019}), \bibinfo{pages}{445--458}.
\newblock


\bibitem[Aceto et~al\mbox{.}(2019b)]%
        {aceto2019mimetic}
\bibfield{author}{\bibinfo{person}{Giuseppe Aceto}, \bibinfo{person}{Domenico Ciuonzo}, \bibinfo{person}{Antonio Montieri}, {and} \bibinfo{person}{Antonio Pescap{\`e}}.} \bibinfo{year}{2019}\natexlab{b}.
\newblock \showarticletitle{MIMETIC: Mobile encrypted traffic classification using multimodal deep learning}. In \bibinfo{booktitle}{\emph{Computer networks}}, Vol.~\bibinfo{volume}{165}. \bibinfo{publisher}{Elsevier}, \bibinfo{pages}{106944}.
\newblock


\bibitem[Aceto et~al\mbox{.}(2020)]%
        {aceto2020toward}
\bibfield{author}{\bibinfo{person}{Giuseppe Aceto}, \bibinfo{person}{Domenico Ciuonzo}, \bibinfo{person}{Antonio Montieri}, {and} \bibinfo{person}{Antonio Pescap{\'e}}.} \bibinfo{year}{2020}\natexlab{}.
\newblock \showarticletitle{Toward effective mobile encrypted traffic classification through deep learning}.
\newblock \bibinfo{journal}{\emph{Neurocomputing}}  \bibinfo{volume}{409} (\bibinfo{year}{2020}), \bibinfo{pages}{306--315}.
\newblock


\bibitem[Akbari et~al\mbox{.}(2023)]%
        {elham_cnsm}
\bibfield{author}{\bibinfo{person}{Elham Akbari}, \bibinfo{person}{Sheikh~A. Tahmid}, \bibinfo{person}{Navid Malekghaeini}, \bibinfo{person}{M.A. Salahuddin}, \bibinfo{person}{Noura Limam}, \bibinfo{person}{Raouf Boutaba}, {et~al\mbox{.}}} \bibinfo{year}{2023}\natexlab{}.
\newblock \showarticletitle{A {C}ritical {S}tudy of {F}ew-shot {L}earning for {E}ncrypted {T}raffic {C}lassification}. In \bibinfo{booktitle}{\emph{Proceedings of 19th International Conference on Service and Network Management}}.
\newblock


\bibitem[Akbari et~al\mbox{.}(2021)]%
        {iman}
\bibfield{author}{\bibinfo{person}{Iman Akbari}, \bibinfo{person}{Mohammad~A Salahuddin}, \bibinfo{person}{Leni Ven}, \bibinfo{person}{Noura Limam}, \bibinfo{person}{Raouf Boutaba}, \bibinfo{person}{Bertrand Mathieu}, \bibinfo{person}{Stephanie Moteau}, {and} \bibinfo{person}{Stephane Tuffin}.} \bibinfo{year}{2021}\natexlab{}.
\newblock \showarticletitle{A look behind the curtain: traffic classification in an increasingly encrypted web}.
\newblock \bibinfo{journal}{\emph{Proceedings of the ACM on Measurement and Analysis of Computing Systems}} \bibinfo{volume}{5}, \bibinfo{number}{1} (\bibinfo{year}{2021}), \bibinfo{pages}{1--26}.
\newblock


\bibitem[Anderson and McGrew(2017)]%
        {randomforestforTC2017}
\bibfield{author}{\bibinfo{person}{Blake Anderson} {and} \bibinfo{person}{David McGrew}.} \bibinfo{year}{2017}\natexlab{}.
\newblock \showarticletitle{Machine learning for encrypted malware traffic classification: accounting for noisy labels and non-stationarity}. In \bibinfo{booktitle}{\emph{Proceedings of the 23rd ACM SIGKDD International Conference on knowledge discovery and data mining}}. \bibinfo{pages}{1723--1732}.
\newblock


\bibitem[Arjovsky et~al\mbox{.}(2019)]%
        {arjovsky2019invariant}
\bibfield{author}{\bibinfo{person}{Martin Arjovsky}, \bibinfo{person}{L{\'e}on Bottou}, \bibinfo{person}{Ishaan Gulrajani}, {and} \bibinfo{person}{David Lopez-Paz}.} \bibinfo{year}{2019}\natexlab{}.
\newblock \showarticletitle{Invariant risk minimization}.
\newblock \bibinfo{journal}{\emph{arXiv preprint arXiv:1907.02893}} (\bibinfo{year}{2019}).
\newblock


\bibitem[Arp et~al\mbox{.}(2022)]%
        {dosanddonts22}
\bibfield{author}{\bibinfo{person}{Daniel Arp}, \bibinfo{person}{Erwin Quiring}, \bibinfo{person}{Feargus Pendlebury}, \bibinfo{person}{Alexander Warnecke}, \bibinfo{person}{Fabio Pierazzi}, \bibinfo{person}{Christian Wressnegger}, \bibinfo{person}{Lorenzo Cavallaro}, {and} \bibinfo{person}{Konrad Rieck}.} \bibinfo{year}{2022}\natexlab{}.
\newblock \showarticletitle{Dos and don'ts of machine learning in computer security}. In \bibinfo{booktitle}{\emph{31st USENIX Security Symposium (USENIX Security 22)}}. \bibinfo{pages}{3971--3988}.
\newblock


\bibitem[Beltiukov et~al\mbox{.}(2023)]%
        {netunicorn_ccs2023}
\bibfield{author}{\bibinfo{person}{Roman Beltiukov}, \bibinfo{person}{Wenbo Guo}, \bibinfo{person}{Arpit Gupta}, {and} \bibinfo{person}{Walter Willinger}.} \bibinfo{year}{2023}\natexlab{}.
\newblock \showarticletitle{In Search of netUnicorn: A Data-Collection Platform to Develop Generalizable ML Models for Network Security Problems}. In \bibinfo{booktitle}{\emph{Proceedings of the 2023 ACM SIGSAC Conference on Computer and Communications Security}}. \bibinfo{pages}{2217--2231}.
\newblock


\bibitem[Bronzino et~al\mbox{.}(2019)]%
        {bronzino2019inferring}
\bibfield{author}{\bibinfo{person}{Francesco Bronzino}, \bibinfo{person}{Paul Schmitt}, \bibinfo{person}{Sara Ayoubi}, \bibinfo{person}{Guilherme Martins}, \bibinfo{person}{Renata Teixeira}, {and} \bibinfo{person}{Nick Feamster}.} \bibinfo{year}{2019}\natexlab{}.
\newblock \showarticletitle{Inferring streaming video quality from encrypted traffic: Practical models and deployment experience}.
\newblock \bibinfo{journal}{\emph{Proceedings of the ACM on Measurement and Analysis of Computing Systems}} \bibinfo{volume}{3}, \bibinfo{number}{3} (\bibinfo{year}{2019}), \bibinfo{pages}{1--25}.
\newblock


\bibitem[Chen and Guestrin(2016)]%
        {xgboost2016first}
\bibfield{author}{\bibinfo{person}{Tianqi Chen} {and} \bibinfo{person}{Carlos Guestrin}.} \bibinfo{year}{2016}\natexlab{}.
\newblock \showarticletitle{Xgboost: A scalable tree boosting system}.
\newblock  (\bibinfo{year}{2016}), \bibinfo{pages}{785--794}.
\newblock


\bibitem[D'Amour et~al\mbox{.}(2022)]%
        {ml_underspecification22}
\bibfield{author}{\bibinfo{person}{Alexander D'Amour}, \bibinfo{person}{Katherine Heller}, \bibinfo{person}{Dan Moldovan}, \bibinfo{person}{Ben Adlam}, \bibinfo{person}{Babak Alipanahi}, \bibinfo{person}{Alex Beutel}, \bibinfo{person}{Christina Chen}, \bibinfo{person}{Jonathan Deaton}, \bibinfo{person}{Jacob Eisenstein}, \bibinfo{person}{Matthew~D Hoffman}, {et~al\mbox{.}}} \bibinfo{year}{2022}\natexlab{}.
\newblock \showarticletitle{Underspecification presents challenges for credibility in modern machine learning}.
\newblock \bibinfo{journal}{\emph{The Journal of Machine Learning Research}} \bibinfo{volume}{23}, \bibinfo{number}{1} (\bibinfo{year}{2022}), \bibinfo{pages}{10237--10297}.
\newblock


\bibitem[David et~al\mbox{.}(2010)]%
        {bendavid2010domainadapt}
\bibfield{author}{\bibinfo{person}{Shai~Ben David}, \bibinfo{person}{Tyler Lu}, \bibinfo{person}{Teresa Luu}, {and} \bibinfo{person}{D{\'a}vid P{\'a}l}.} \bibinfo{year}{2010}\natexlab{}.
\newblock \showarticletitle{Impossibility theorems for domain adaptation}. In \bibinfo{booktitle}{\emph{Proceedings of the Thirteenth International Conference on Artificial Intelligence and Statistics}}. JMLR Workshop and Conference Proceedings, \bibinfo{pages}{129--136}.
\newblock


\bibitem[Deng et~al\mbox{.}(2023)]%
        {wfattacks2023}
\bibfield{author}{\bibinfo{person}{Xinhao Deng}, \bibinfo{person}{Qilei Yin}, \bibinfo{person}{Zhuotao Liu}, \bibinfo{person}{Xiyuan Zhao}, \bibinfo{person}{Qi Li}, \bibinfo{person}{Mingwei Xu}, \bibinfo{person}{Ke Xu}, {and} \bibinfo{person}{Jianping Wu}.} \bibinfo{year}{2023}\natexlab{}.
\newblock \showarticletitle{Robust Multi-tab Website Fingerprinting Attacks in the Wild}. In \bibinfo{booktitle}{\emph{2023 IEEE Symposium on Security and Privacy (SP)}}. \bibinfo{pages}{1005--1022}.
\newblock
\href{https://doi.org/10.1109/SP46215.2023.10179464}{doi:\nolinkurl{10.1109/SP46215.2023.10179464}}


\bibitem[Draper-Gil et~al\mbox{.}(2016)]%
        {iscx-vpn}
\bibfield{author}{\bibinfo{person}{Gerard Draper-Gil}, \bibinfo{person}{Arash~Habibi Lashkari}, \bibinfo{person}{Mohammad Saiful~Islam Mamun}, {and} \bibinfo{person}{Ali~A Ghorbani}.} \bibinfo{year}{2016}\natexlab{}.
\newblock \showarticletitle{Characterization of encrypted and vpn traffic using time-related}. In \bibinfo{booktitle}{\emph{Proceedings of the 2nd international conference on information systems security and privacy (ICISSP)}}. \bibinfo{pages}{407--414}.
\newblock


\bibitem[Fang et~al\mbox{.}(2020)]%
        {dl_distributionshift2020}
\bibfield{author}{\bibinfo{person}{Tongtong Fang}, \bibinfo{person}{Nan Lu}, \bibinfo{person}{Gang Niu}, {and} \bibinfo{person}{Masashi Sugiyama}.} \bibinfo{year}{2020}\natexlab{}.
\newblock \showarticletitle{Rethinking importance weighting for deep learning under distribution shift}.
\newblock \bibinfo{journal}{\emph{Advances in neural information processing systems}}  \bibinfo{volume}{33} (\bibinfo{year}{2020}), \bibinfo{pages}{11996--12007}.
\newblock


\bibitem[Finamore et~al\mbox{.}(2023)]%
        {flowpic2023replication}
\bibfield{author}{\bibinfo{person}{Alessandro Finamore}, \bibinfo{person}{Chao Wang}, \bibinfo{person}{Jonatan Krolikowski}, \bibinfo{person}{Jose~M Navarro}, \bibinfo{person}{Fuxing Chen}, {and} \bibinfo{person}{Dario Rossi}.} \bibinfo{year}{2023}\natexlab{}.
\newblock \showarticletitle{Replication: Contrastive learning and data augmentation in traffic classification using a Flowpic input representation}. In \bibinfo{booktitle}{\emph{Proceedings of the 2023 ACM on internet measurement conference}}. \bibinfo{pages}{36--51}.
\newblock


\bibitem[Foret et~al\mbox{.}(2020)]%
        {neyshabur2020sharpness}
\bibfield{author}{\bibinfo{person}{Pierre Foret}, \bibinfo{person}{Ariel Kleiner}, \bibinfo{person}{Hossein Mobahi}, {and} \bibinfo{person}{Behnam Neyshabur}.} \bibinfo{year}{2020}\natexlab{}.
\newblock \showarticletitle{Sharpness-aware minimization for efficiently improving generalization}.
\newblock \bibinfo{journal}{\emph{arXiv preprint arXiv:2010.01412}} (\bibinfo{year}{2020}).
\newblock


\bibitem[Garg et~al\mbox{.}(2022)]%
        {neyshabur2022unlabeled-ood}
\bibfield{author}{\bibinfo{person}{Saurabh Garg}, \bibinfo{person}{Sivaraman Balakrishnan}, \bibinfo{person}{Zachary~C Lipton}, \bibinfo{person}{Behnam Neyshabur}, {and} \bibinfo{person}{Hanie Sedghi}.} \bibinfo{year}{2022}\natexlab{}.
\newblock \showarticletitle{Leveraging unlabeled data to predict out-of-distribution performance}.
\newblock \bibinfo{journal}{\emph{arXiv preprint arXiv:2201.04234}} (\bibinfo{year}{2022}).
\newblock


\bibitem[Geirhos et~al\mbox{.}(2020)]%
        {nature_gen20}
\bibfield{author}{\bibinfo{person}{Robert Geirhos}, \bibinfo{person}{J{\"o}rn-Henrik Jacobsen}, \bibinfo{person}{Claudio Michaelis}, \bibinfo{person}{Richard Zemel}, \bibinfo{person}{Wieland Brendel}, \bibinfo{person}{Matthias Bethge}, {and} \bibinfo{person}{Felix~A Wichmann}.} \bibinfo{year}{2020}\natexlab{}.
\newblock \showarticletitle{Shortcut learning in deep neural networks}.
\newblock \bibinfo{journal}{\emph{Nature Machine Intelligence}} \bibinfo{volume}{2}, \bibinfo{number}{11} (\bibinfo{year}{2020}), \bibinfo{pages}{665--673}.
\newblock


\bibitem[Guo et~al\mbox{.}(2022)]%
        {guo2022ifip}
\bibfield{author}{\bibinfo{person}{Jingyu Guo} {et~al\mbox{.}}} \bibinfo{year}{2022}\natexlab{}.
\newblock \showarticletitle{Global-Aware Prototypical Network for Few-Shot Encrypted Traffic Classification}. In \bibinfo{booktitle}{\emph{2022 IFIP Networking Conference (IFIP Networking)}}. IEEE, \bibinfo{pages}{1--9}.
\newblock


\bibitem[Guthula et~al\mbox{.}(2023)]%
        {netfound23}
\bibfield{author}{\bibinfo{person}{Satyandra Guthula}, \bibinfo{person}{Navya Battula}, \bibinfo{person}{Roman Beltiukov}, \bibinfo{person}{Wenbo Guo}, {and} \bibinfo{person}{Arpit Gupta}.} \bibinfo{year}{2023}\natexlab{}.
\newblock \showarticletitle{netFound: Foundation Model for Network Security}.
\newblock \bibinfo{journal}{\emph{arXiv preprint arXiv:2310.17025}} (\bibinfo{year}{2023}).
\newblock


\bibitem[Herrmann et~al\mbox{.}(2009)]%
        {wf2009herrmann}
\bibfield{author}{\bibinfo{person}{Dominik Herrmann}, \bibinfo{person}{Rolf Wendolsky}, {and} \bibinfo{person}{Hannes Federrath}.} \bibinfo{year}{2009}\natexlab{}.
\newblock \showarticletitle{Website fingerprinting: attacking popular privacy enhancing technologies with the multinomial na{\"\i}ve-bayes classifier}. In \bibinfo{booktitle}{\emph{Proceedings of the 2009 ACM workshop on Cloud computing security}}.
\newblock


\bibitem[Hoang et~al\mbox{.}(2020)]%
        {esni2020assessing}
\bibfield{author}{\bibinfo{person}{Nguyen~Phong Hoang}, \bibinfo{person}{Arian Akhavan~Niaki}, \bibinfo{person}{Nikita Borisov}, \bibinfo{person}{Phillipa Gill}, {and} \bibinfo{person}{Michalis Polychronakis}.} \bibinfo{year}{2020}\natexlab{}.
\newblock \showarticletitle{Assessing the privacy benefits of domain name encryption}. In \bibinfo{booktitle}{\emph{Proceedings of the 15th ACM Asia Conference on Computer and Communications Security}}. \bibinfo{pages}{290--304}.
\newblock


\bibitem[Holland et~al\mbox{.}(2021)]%
        {nprint}
\bibfield{author}{\bibinfo{person}{Jordan Holland}, \bibinfo{person}{Paul Schmitt}, \bibinfo{person}{Nick Feamster}, {and} \bibinfo{person}{Prateek Mittal}.} \bibinfo{year}{2021}\natexlab{}.
\newblock \showarticletitle{New directions in automated traffic analysis}. In \bibinfo{booktitle}{\emph{Proceedings of the 2021 ACM SIGSAC Conference on Computer and Communications Security}}. \bibinfo{pages}{3366--3383}.
\newblock


\bibitem[Horowicz et~al\mbox{.}(2022)]%
        {miniflowpic}
\bibfield{author}{\bibinfo{person}{Eyal Horowicz}, \bibinfo{person}{Tal Shapira}, {and} \bibinfo{person}{Yuval Shavitt}.} \bibinfo{year}{2022}\natexlab{}.
\newblock \showarticletitle{A few shots traffic classification with mini-FlowPic augmentations}. In \bibinfo{booktitle}{\emph{Proceedings of the 22nd ACM Internet Measurement Conference}}. \bibinfo{pages}{647--654}.
\newblock


\bibitem[Huoh et~al\mbox{.}(2023)]%
        {gnn2023tnsm}
\bibfield{author}{\bibinfo{person}{Ting-Li Huoh}, \bibinfo{person}{Yan Luo}, \bibinfo{person}{Peilong Li}, {and} \bibinfo{person}{Tong Zhang}.} \bibinfo{year}{2023}\natexlab{}.
\newblock \showarticletitle{Flow-Based Encrypted Network Traffic Classification With Graph Neural Networks}.
\newblock \bibinfo{journal}{\emph{IEEE Transactions on Network and Service Management}} \bibinfo{volume}{20}, \bibinfo{number}{2} (\bibinfo{year}{2023}), \bibinfo{pages}{1224--1237}.
\newblock
\href{https://doi.org/10.1109/TNSM.2022.3227500}{doi:\nolinkurl{10.1109/TNSM.2022.3227500}}


\bibitem[Ilbert et~al\mbox{.}(2024)]%
        {samformer2024}
\bibfield{author}{\bibinfo{person}{Romain Ilbert}, \bibinfo{person}{Ambroise Odonnat}, \bibinfo{person}{Vasilii Feofanov}, \bibinfo{person}{Aladin Virmaux}, \bibinfo{person}{Giuseppe Paolo}, \bibinfo{person}{Themis Palpanas}, {and} \bibinfo{person}{Ievgen Redko}.} \bibinfo{year}{2024}\natexlab{}.
\newblock \showarticletitle{SAMformer: unlocking the potential of transformers in time series forecasting with sharpness-aware minimization and channel-wise attention}.
\newblock , Article \bibinfo{articleno}{841} (\bibinfo{year}{2024}), \bibinfo{numpages}{31}~pages.
\newblock


\bibitem[Inacio and Trammell(2010)]%
        {yaf}
\bibfield{author}{\bibinfo{person}{Christopher~M Inacio} {and} \bibinfo{person}{Brian Trammell}.} \bibinfo{year}{2010}\natexlab{}.
\newblock \showarticletitle{$\{$YAF$\}$: Yet Another Flowmeter}. In \bibinfo{booktitle}{\emph{24th Large Installation System Administration Conference (LISA 10)}}.
\newblock


\bibitem[Jacobs et~al\mbox{.}(2022)]%
        {trustee}
\bibfield{author}{\bibinfo{person}{Arthur~S Jacobs}, \bibinfo{person}{Roman Beltiukov}, \bibinfo{person}{Walter Willinger}, \bibinfo{person}{Ronaldo~A Ferreira}, \bibinfo{person}{Arpit Gupta}, {and} \bibinfo{person}{Lisandro~Z Granville}.} \bibinfo{year}{2022}\natexlab{}.
\newblock \showarticletitle{Ai/ml for network security: The emperor has no clothes}. In \bibinfo{booktitle}{\emph{Proceedings of the 2022 ACM SIGSAC Conference on Computer and Communications Security}}. \bibinfo{pages}{1537--1551}.
\newblock


\bibitem[Jiang et~al\mbox{.}(2019)]%
        {generalization2019neyshabur}
\bibfield{author}{\bibinfo{person}{Yiding Jiang}, \bibinfo{person}{Behnam Neyshabur}, \bibinfo{person}{Hossein Mobahi}, \bibinfo{person}{Dilip Krishnan}, {and} \bibinfo{person}{Samy Bengio}.} \bibinfo{year}{2019}\natexlab{}.
\newblock \showarticletitle{Fantastic generalization measures and where to find them}.
\newblock \bibinfo{journal}{\emph{arXiv preprint arXiv:1912.02178}} (\bibinfo{year}{2019}).
\newblock


\bibitem[Juarez et~al\mbox{.}(2014)]%
        {wfcriticaleval2014}
\bibfield{author}{\bibinfo{person}{Marc Juarez}, \bibinfo{person}{Sadia Afroz}, \bibinfo{person}{Gunes Acar}, \bibinfo{person}{Claudia Diaz}, {and} \bibinfo{person}{Rachel Greenstadt}.} \bibinfo{year}{2014}\natexlab{}.
\newblock \showarticletitle{A critical evaluation of website fingerprinting attacks}. In \bibinfo{booktitle}{\emph{Proceedings of the 2014 ACM SIGSAC Conference on Computer and Communications Security}}. \bibinfo{pages}{263--274}.
\newblock


\bibitem[Kingma and Ba(2014)]%
        {adam}
\bibfield{author}{\bibinfo{person}{Diederik~P Kingma} {and} \bibinfo{person}{Jimmy Ba}.} \bibinfo{year}{2014}\natexlab{}.
\newblock \showarticletitle{Adam: A method for stochastic optimization}.
\newblock \bibinfo{journal}{\emph{arXiv preprint arXiv:1412.6980}} (\bibinfo{year}{2014}).
\newblock


\bibitem[Koh et~al\mbox{.}(2021)]%
        {saca2021wilds}
\bibfield{author}{\bibinfo{person}{Pang~Wei Koh}, \bibinfo{person}{Shiori Sagawa}, \bibinfo{person}{Henrik Marklund}, \bibinfo{person}{Sang~Michael Xie}, \bibinfo{person}{Marvin Zhang}, \bibinfo{person}{Akshay Balsubramani}, \bibinfo{person}{Weihua Hu}, \bibinfo{person}{Michihiro Yasunaga}, \bibinfo{person}{Richard~Lanas Phillips}, \bibinfo{person}{Irena Gao}, {et~al\mbox{.}}} \bibinfo{year}{2021}\natexlab{}.
\newblock \showarticletitle{Wilds: A benchmark of in-the-wild distribution shifts}. In \bibinfo{booktitle}{\emph{International conference on machine learning}}. PMLR, \bibinfo{pages}{5637--5664}.
\newblock


\bibitem[Kosson et~al\mbox{.}(2024)]%
        {rotational2024equilibrium}
\bibfield{author}{\bibinfo{person}{Atli Kosson}, \bibinfo{person}{Bettina Messmer}, {and} \bibinfo{person}{Martin Jaggi}.} \bibinfo{year}{2024}\natexlab{}.
\newblock \showarticletitle{Rotational equilibrium: how weight decay balances learning across neural networks}.
\newblock , Article \bibinfo{articleno}{1015} (\bibinfo{year}{2024}), \bibinfo{numpages}{37}~pages.
\newblock


\bibitem[Lashkari et~al\mbox{.}(2017)]%
        {iscx-tor}
\bibfield{author}{\bibinfo{person}{Arash~Habibi Lashkari}, \bibinfo{person}{Gerard~Draper Gil}, \bibinfo{person}{Mohammad Saiful~Islam Mamun}, {and} \bibinfo{person}{Ali~A Ghorbani}.} \bibinfo{year}{2017}\natexlab{}.
\newblock \showarticletitle{Characterization of tor traffic using time based features}. In \bibinfo{booktitle}{\emph{International Conference on Information Systems Security and Privacy}}, Vol.~\bibinfo{volume}{2}. SciTePress, \bibinfo{pages}{253--262}.
\newblock


\bibitem[Li et~al\mbox{.}(2022)]%
        {appfingerprinting2022}
\bibfield{author}{\bibinfo{person}{Jianfeng Li}, \bibinfo{person}{Hao Zhou}, \bibinfo{person}{Shuohan Wu}, \bibinfo{person}{Xiapu Luo}, \bibinfo{person}{Ting Wang}, \bibinfo{person}{Xian Zhan}, {and} \bibinfo{person}{Xiaobo Ma}.} \bibinfo{year}{2022}\natexlab{}.
\newblock \showarticletitle{$\{$FOAP$\}$:$\{$Fine-Grained$\}$$\{$Open-World$\}$ android app fingerprinting}. In \bibinfo{booktitle}{\emph{31st USENIX Security Symposium (USENIX Security 22)}}. \bibinfo{pages}{1579--1596}.
\newblock


\bibitem[Lin et~al\mbox{.}(2022)]%
        {etbert}
\bibfield{author}{\bibinfo{person}{Xinjie Lin} {et~al\mbox{.}}} \bibinfo{year}{2022}\natexlab{}.
\newblock \showarticletitle{Et-bert: A contextualized datagram representation with pre-training transformers for encrypted traffic classification}. In \bibinfo{booktitle}{\emph{Proceedings of the ACM Web Conference 2022}}. \bibinfo{pages}{633--642}.
\newblock


\bibitem[Liu et~al\mbox{.}(2020)]%
        {liu2020trainingdifficulty}
\bibfield{author}{\bibinfo{person}{Liyuan Liu}, \bibinfo{person}{Xiaodong Liu}, \bibinfo{person}{Jianfeng Gao}, \bibinfo{person}{Weizhu Chen}, {and} \bibinfo{person}{Jiawei Han}.} \bibinfo{year}{2020}\natexlab{}.
\newblock \showarticletitle{Understanding the difficulty of training transformers}.
\newblock \bibinfo{journal}{\emph{arXiv preprint arXiv:2004.08249}} (\bibinfo{year}{2020}).
\newblock


\bibitem[Luxemburk and Čejka(2023)]%
        {cesnet}
\bibfield{author}{\bibinfo{person}{Jan Luxemburk} {and} \bibinfo{person}{Tomáš Čejka}.} \bibinfo{year}{2023}\natexlab{}.
\newblock \showarticletitle{Fine-grained TLS services classification with reject option}.
\newblock \bibinfo{journal}{\emph{Computer Networks}}  \bibinfo{volume}{220} (\bibinfo{year}{2023}), \bibinfo{pages}{109467}.
\newblock
\showISSN{1389-1286}
\href{https://doi.org/10.1016/j.comnet.2022.109467}{doi:\nolinkurl{10.1016/j.comnet.2022.109467}}


\bibitem[Malekghaini et~al\mbox{.}(2022)]%
        {navid2022datadrift}
\bibfield{author}{\bibinfo{person}{Navid Malekghaini}, \bibinfo{person}{Elham Akbari}, \bibinfo{person}{Mohammad~A Salahuddin}, \bibinfo{person}{Noura Limam}, \bibinfo{person}{Raouf Boutaba}, \bibinfo{person}{Bertrand Mathieu}, \bibinfo{person}{Stephanie Moteau}, {and} \bibinfo{person}{Stephane Tuffin}.} \bibinfo{year}{2022}\natexlab{}.
\newblock \showarticletitle{Data Drift in DL: Lessons Learned from Encrypted Traffic Classification}. In \bibinfo{booktitle}{\emph{2022 IFIP Networking Conference (IFIP Networking)}}. IEEE, \bibinfo{pages}{1--9}.
\newblock


\bibitem[Nascita et~al\mbox{.}(2021)]%
        {pescape2021xaimirage}
\bibfield{author}{\bibinfo{person}{Alfredo Nascita}, \bibinfo{person}{Antonio Montieri}, \bibinfo{person}{Giuseppe Aceto}, \bibinfo{person}{Domenico Ciuonzo}, \bibinfo{person}{Valerio Persico}, {and} \bibinfo{person}{Antonio Pescap{\'e}}.} \bibinfo{year}{2021}\natexlab{}.
\newblock \showarticletitle{XAI meets mobile traffic classification: Understanding and improving multimodal deep learning architectures}.
\newblock \bibinfo{journal}{\emph{IEEE Transactions on Network and Service Management}} \bibinfo{volume}{18}, \bibinfo{number}{4} (\bibinfo{year}{2021}), \bibinfo{pages}{4225--4246}.
\newblock


\bibitem[Neyshabur et~al\mbox{.}(2020)]%
        {neyshabur2020transfer}
\bibfield{author}{\bibinfo{person}{Behnam Neyshabur}, \bibinfo{person}{Hanie Sedghi}, {and} \bibinfo{person}{Chiyuan Zhang}.} \bibinfo{year}{2020}\natexlab{}.
\newblock \showarticletitle{What is being transferred in transfer learning?}
\newblock \bibinfo{journal}{\emph{Advances in neural information processing systems}}  \bibinfo{volume}{33} (\bibinfo{year}{2020}), \bibinfo{pages}{512--523}.
\newblock


\bibitem[Nie et~al\mbox{.}(2023)]%
        {worth64words2022}
\bibfield{author}{\bibinfo{person}{Yuqi Nie}, \bibinfo{person}{Nam~H Nguyen}, \bibinfo{person}{Phanwadee Sinthong}, {and} \bibinfo{person}{Jayant Kalagnanam}.} \bibinfo{year}{2023}\natexlab{}.
\newblock \showarticletitle{A Time Series is Worth 64 Words: Long-term Forecasting with Transformers}. In \bibinfo{booktitle}{\emph{The Eleventh International Conference on Learning Representations}}.
\newblock
\urldef\tempurl%
\url{https://openreview.net/forum?id=Jbdc0vTOcol}
\showURL{%
\tempurl}


\bibitem[Piet et~al\mbox{.}(2023)]%
        {ggfast}
\bibfield{author}{\bibinfo{person}{Julien Piet}, \bibinfo{person}{Dubem Nwoji}, {and} \bibinfo{person}{Vern Paxson}.} \bibinfo{year}{2023}\natexlab{}.
\newblock \showarticletitle{{GGFAST}: Automating Generation of Flexible Network Traffic Classifiers}. In \bibinfo{booktitle}{\emph{Proceedings of the ACM SIGCOMM 2023 Conference}}. \bibinfo{pages}{850--866}.
\newblock


\bibitem[Rahman et~al\mbox{.}(2019)]%
        {tiktok2019}
\bibfield{author}{\bibinfo{person}{Mohammad~Saidur Rahman}, \bibinfo{person}{Payap Sirinam}, \bibinfo{person}{Nate Mathews}, \bibinfo{person}{Kantha~Girish Gangadhara}, {and} \bibinfo{person}{Matthew Wright}.} \bibinfo{year}{2019}\natexlab{}.
\newblock \showarticletitle{Tik-tok: The utility of packet timing in website fingerprinting attacks}.
\newblock \bibinfo{journal}{\emph{arXiv preprint arXiv:1902.06421}} (\bibinfo{year}{2019}).
\newblock


\bibitem[Shapira and Shavitt(2019)]%
        {flowpic}
\bibfield{author}{\bibinfo{person}{Tal Shapira} {and} \bibinfo{person}{Yuval Shavitt}.} \bibinfo{year}{2019}\natexlab{}.
\newblock \showarticletitle{Flowpic: Encrypted internet traffic classification is as easy as image recognition}. In \bibinfo{booktitle}{\emph{IEEE INFOCOM 2019-IEEE Conference on Computer Communications Workshops (INFOCOM WKSHPS)}}. IEEE, \bibinfo{pages}{680--687}.
\newblock


\bibitem[Sirinam et~al\mbox{.}(2019)]%
        {wftriplet}
\bibfield{author}{\bibinfo{person}{Payap Sirinam} {et~al\mbox{.}}} \bibinfo{year}{2019}\natexlab{}.
\newblock \showarticletitle{Triplet fingerprinting: More practical and portable website fingerprinting with n-shot learning}. In \bibinfo{booktitle}{\emph{Proceedings of the 2019 ACM SIGSAC Conference on Computer and Communications Security}}. \bibinfo{pages}{1131--1148}.
\newblock


\bibitem[Sirinam et~al\mbox{.}(2018)]%
        {dfattack}
\bibfield{author}{\bibinfo{person}{Payap Sirinam}, \bibinfo{person}{Mohsen Imani}, \bibinfo{person}{Marc Juarez}, {and} \bibinfo{person}{Matthew Wright}.} \bibinfo{year}{2018}\natexlab{}.
\newblock \showarticletitle{Deep fingerprinting: Undermining website fingerprinting defenses with deep learning}. In \bibinfo{booktitle}{\emph{Proceedings of the 2018 ACM SIGSAC conference on computer and communications security}}. \bibinfo{pages}{1928--1943}.
\newblock


\bibitem[Smith(2017)]%
        {cyclicLR}
\bibfield{author}{\bibinfo{person}{Leslie~N. Smith}.} \bibinfo{year}{2017}\natexlab{}.
\newblock \showarticletitle{Cyclical Learning Rates for Training Neural Networks}. In \bibinfo{booktitle}{\emph{2017 IEEE Winter Conference on Applications of Computer Vision (WACV)}}. \bibinfo{pages}{464--472}.
\newblock
\href{https://doi.org/10.1109/WACV.2017.58}{doi:\nolinkurl{10.1109/WACV.2017.58}}


\bibitem[Taylor et~al\mbox{.}(2016)]%
        {taylor2016appscanner}
\bibfield{author}{\bibinfo{person}{Vincent~F Taylor}, \bibinfo{person}{Riccardo Spolaor}, \bibinfo{person}{Mauro Conti}, {and} \bibinfo{person}{Ivan Martinovic}.} \bibinfo{year}{2016}\natexlab{}.
\newblock \showarticletitle{Appscanner: Automatic fingerprinting of smartphone apps from encrypted network traffic}. In \bibinfo{booktitle}{\emph{2016 IEEE European Symposium on Security and Privacy (EuroS\&P)}}. IEEE, \bibinfo{pages}{439--454}.
\newblock


\bibitem[Van~Ede et~al\mbox{.}(2020)]%
        {flowprint2020}
\bibfield{author}{\bibinfo{person}{Thijs Van~Ede}, \bibinfo{person}{Riccardo Bortolameotti}, \bibinfo{person}{Andrea Continella}, \bibinfo{person}{Jingjing Ren}, \bibinfo{person}{Daniel~J Dubois}, \bibinfo{person}{Martina Lindorfer}, \bibinfo{person}{David Choffnes}, \bibinfo{person}{Maarten van Steen}, {and} \bibinfo{person}{Andreas Peter}.} \bibinfo{year}{2020}\natexlab{}.
\newblock \showarticletitle{Flowprint: Semi-supervised mobile-app fingerprinting on encrypted network traffic}. In \bibinfo{booktitle}{\emph{Network and distributed system security symposium (NDSS)}}, Vol.~\bibinfo{volume}{27}.
\newblock


\bibitem[Wang and Goldberg(2016)]%
        {wfrealistictorattack2016}
\bibfield{author}{\bibinfo{person}{Tao Wang} {and} \bibinfo{person}{Ian Goldberg}.} \bibinfo{year}{2016}\natexlab{}.
\newblock \showarticletitle{On Realistically Attacking Tor with Website Fingerprinting.}
\newblock \bibinfo{journal}{\emph{Proc. Priv. Enhancing Technol.}} \bibinfo{volume}{2016}, \bibinfo{number}{4} (\bibinfo{year}{2016}), \bibinfo{pages}{21--36}.
\newblock


\bibitem[Wang et~al\mbox{.}(2017)]%
        {wang20171dcnn}
\bibfield{author}{\bibinfo{person}{Wei Wang}, \bibinfo{person}{Ming Zhu}, \bibinfo{person}{Jinlin Wang}, \bibinfo{person}{Xuewen Zeng}, {and} \bibinfo{person}{Zhongzhen Yang}.} \bibinfo{year}{2017}\natexlab{}.
\newblock \showarticletitle{End-to-end encrypted traffic classification with one-dimensional convolution neural networks}. In \bibinfo{booktitle}{\emph{2017 IEEE international conference on intelligence and security informatics (ISI)}}. IEEE, \bibinfo{pages}{43--48}.
\newblock


\bibitem[Yang et~al\mbox{.}(2021)]%
        {rossilessonslearned}
\bibfield{author}{\bibinfo{person}{Lixuan Yang}, \bibinfo{person}{Alessandro Finamore}, \bibinfo{person}{Feng Jun}, {and} \bibinfo{person}{Dario Rossi}.} \bibinfo{year}{2021}\natexlab{}.
\newblock \showarticletitle{Deep learning and zero-day traffic classification: Lessons learned from a commercial-grade dataset}. In \bibinfo{booktitle}{\emph{IEEE Transactions on Network and Service Management}}, Vol.~\bibinfo{volume}{18}. \bibinfo{publisher}{IEEE}, \bibinfo{pages}{4103--4118}.
\newblock


\bibitem[Zeng et~al\mbox{.}(2023)]%
        {zeng2023transformers}
\bibfield{author}{\bibinfo{person}{Ailing Zeng}, \bibinfo{person}{Muxi Chen}, \bibinfo{person}{Lei Zhang}, {and} \bibinfo{person}{Qiang Xu}.} \bibinfo{year}{2023}\natexlab{}.
\newblock \showarticletitle{Are transformers effective for time series forecasting?}. In \bibinfo{booktitle}{\emph{Proceedings of the AAAI conference on artificial intelligence}}, Vol.~\bibinfo{volume}{37}. \bibinfo{pages}{11121--11128}.
\newblock


\bibitem[Zhao et~al\mbox{.}(2022)]%
        {kdd22flowformer}
\bibfield{author}{\bibinfo{person}{Ruijie Zhao}, \bibinfo{person}{Xianwen Deng}, \bibinfo{person}{Zhicong Yan}, \bibinfo{person}{Jun Ma}, \bibinfo{person}{Zhi Xue}, {and} \bibinfo{person}{Yijun Wang}.} \bibinfo{year}{2022}\natexlab{}.
\newblock \showarticletitle{Mt-flowformer: A semi-supervised flow transformer for encrypted traffic classification}. In \bibinfo{booktitle}{\emph{Proceedings of the 28th ACM SIGKDD Conference on Knowledge Discovery and Data Mining}}. \bibinfo{pages}{2576--2584}.
\newblock


\bibitem[Zhou et~al\mbox{.}(2024)]%
        {trafficformer2024sp}
\bibfield{author}{\bibinfo{person}{Guangmeng Zhou}, \bibinfo{person}{Xiongwen Guo}, \bibinfo{person}{Zhuotao Liu}, \bibinfo{person}{Tong Li}, \bibinfo{person}{Qi Li}, {and} \bibinfo{person}{Ke Xu}.} \bibinfo{year}{2024}\natexlab{}.
\newblock \showarticletitle{Trafficformer: an efficient pre-trained model for traffic data}. In \bibinfo{booktitle}{\emph{2025 IEEE Symposium on Security and Privacy (SP)}}. IEEE Computer Society, \bibinfo{pages}{102--102}.
\newblock


\end{thebibliography}
